\documentclass[10pt,journal,twocolumn,twoside,web]{ieeecolor}
\usepackage{generic}
\usepackage{cite}
\usepackage{amsmath,amssymb,amsfonts}
\usepackage{algorithmic}
\usepackage{graphicx}
\usepackage{textcomp}

\usepackage{lipsum}

\usepackage{tikz}

\usepackage{cite}
\newtheorem{lemma}{\textbf{Lemma}}{}
\newtheorem{corollary}{\textbf{Corollary}}{}
\newtheorem{definition}{\textbf{Definition}}{}

\def\BibTeX{{\rm B\kern-.05em{\sc i\kern-.025em b}\kern-.08em
    T\kern-.1667em\lower.7ex\hbox{E}\kern-.125emX}}
\begin{document}
\title{Stabilizing Error Correction Codes for Control over  Erasure Channels}
\author{Jan {\O}stergaard, \IEEEmembership{Senior Member, IEEE} 
\thanks{J. {\O}stergaard is with the Section on Artificial Intelligence and Sound, Department of Electronic Systems,  Aalborg University, 9220 Aalborg, Denmark.
        {\tt\small jo@es.aau.dk}.}     
\thanks{A brief version of this work has been presented in \cite{ostergaard:2021b}.}
}

\maketitle

\begin{abstract}
We propose $(k,k')$ stabilizing codes, which is a type of delayless error correction codes that are useful for control over networks with erasures.
For each input symbol, $k$ output symbols are generated by the stabilizing code. Receiving at least $k'$ of these outputs guarantees stability. Thus, both the system to be stabilized and the channel is taken into account in the design of the erasure codes. Receiving more than $k'$ outputs further improves the performance of the system. 
In the case of i.i.d.\ erasures, we further demonstrate that the  erasure code can be constructed such that stability is achieved if on average at least $k'$ output symbols are received.
Our focus is on LTI systems, and we construct codes based on independent encodings and multiple descriptions. Stability is assessed via Markov jump linear system theory. The theoretical efficiency and performance of the codes are assessed, and their practical performances are demonstrated in a simulation study. There is a significant gain over other delayless codes such as repetition codes.
\end{abstract}


\section{Introduction}
\label{sec:introduction}
\begin{figure}

\scalebox{0.8}{
\begin{tikzpicture}
\draw   
   (2,3) rectangle (3,2)
	node[pos=0.5]{$P$};

\draw [->] (1.5,2.8) -- (2,2.8)
	node[pos=0.1,above]{$d$};
\draw [->] (1.5,2.5) -- (2,2.5)
	node[pos=0.1,left]{$x_0$};
\draw [->] (0,2.2) -- (2,2.2)
	node[pos=0.1,above]{$u$};
\draw [-] (3,2.2) -- (5,2.2)
	node[pos=0.9,above]{$y$};
\draw [->] (3,2.8) -- (3.5,2.8)
	node[pos=0.9,above]{$e$};

\draw [->] (5,2.2) -- (5,1) -- (4.7,1);
\draw [-] (0,2.2) -- (0,1) -- (0.3,1);

\draw   
   (0.3,0.7) rectangle (1.3,1.3)
	node[pos=0.5]{\footnotesize $\mathcal{D}$}; 

\draw   
   (3.7,0.7) rectangle (4.7,1.3)
	node[pos=0.5]{\footnotesize $\mathcal{E}$};

\draw [<-] (1.3,1) -- (1.9,1);
\draw [<-] (3.1,1) -- (3.7,1);

\draw   
   (1.9,0.7) rectangle (3.1,1.3)
	node[pos=0.5]{\footnotesize Channel};

\end{tikzpicture}
\quad
\raisebox{-5mm}{
\begin{tikzpicture}

\draw   
   (2,3) rectangle (3,2)
	node[pos=0.5]{$P$};

\draw [->] (1.5,2.8) -- (2,2.8)
	node[pos=0.1,above]{$d$};
\draw [->] (1.5,2.5) -- (2,2.5)
	node[pos=0.1,left]{$x_0$};
\draw [->] (0,2.2) -- (2,2.2)
	node[pos=0.1,above]{$u$};
\draw [-] (3,2.2) -- (5,2.2)
	node[pos=0.9,above]{$y$};
\draw [->] (3,2.8) -- (3.5,2.8)
	node[pos=0.9,above]{$e$};

\draw [->] (5,2.2) -- (5,1) -- (4.7,1);
\draw [-] (0,2.2) -- (0,1) -- (0.3,1);

\draw   
   (0.3,0.7) rectangle (1.3,1.3)
	node[pos=0.5]{$F$}; 

\draw   
   (3.7,0.7) rectangle (4.7,1.3)
	node[pos=0.5]{$L$};

\draw [<-] (1.3,1) -- (2.3,1)
	node[pos=0.4,below]{$w$};
\draw [<-] (2.7,1) -- (3.7,1)
	node[pos=0.6,below]{$v$};

\draw [->] (2.51,0.4) -- (2.51,0.8)
	node[pos=0.1,left]{$q$};

\draw   
	(2.1,1.4) rectangle (2.9,1.8)
	node[pos=0.5]{$z^{-1}$}; 

\draw (2.5, 1) circle (6pt)
	node[]{$+$};

\draw [->] (1.6,1) -- (1.6,1.6) -- (2.1,1.6);

\draw [->] (2.9,1.6) -- (4.2,1.6) -- (4.2,1.3);

\end{tikzpicture}}
}
\caption{Left: noisy LTI system $P$ that is controlled over an ideal  digital channel \cite{silva:2016}. Right: linear model of the system.} 
\label{fig:system}
\vspace{-5mm}
\end{figure}
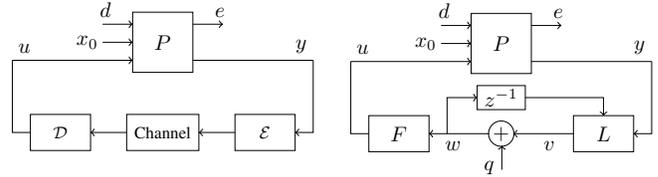

\IEEEPARstart{C}{onsider} 
the networked control system in Fig.~\ref{fig:system}, where the output of a linear and time-invariant  (LTI) open-loop unstable system is to be quantized at a finite data rate and transmitted over a digital channel. For this situation,  a variable-rate and time-varying coding policy was used in \cite{nair:2004} to establish a lower bound on the data rate required to guarantee stability. Stability results have also been obtained for the case of time-invariant fixed-rate quantizers~\cite{yuksel:2010,kostina:2018,sabag:2020} and time-invariant variable-rate quantizers \cite{silva:2011}. The scheme of \cite{silva:2011} was extended beyond stability in \cite{silva:2016} in order to characterize the achievable performance  (measured in terms of the variance of the plant output) that is possible for a given average data rate. If the channel is analogue and Gaussian, it is well-known that linear coding policies are sufficient for stabilizing linear systems~\cite{sahai:2006}. 
For multiple parallel analogue channels, linear coding policies are not necessarily optimal \cite{khina:2019a}.

The situation of estimation or control over channels with erasures have been treated in great detail in the literature, cf.~\cite{tatikonda:,sinopoli:2004,jin:2006,imer:2006,liu:2007,quevedo:2007,ostrovsky:2009,gupta:2009,trivellato:2010,elia:2011,quevedo:2011,garone:2012,yuksel:2013,quevedo:2014,nagahara:2014,2015IJC,ostergaard:2016,peters:2016,maas:2016,khina:2019,barforooshan:2020}, where 
\cite{trivellato:2010,quevedo:2011,yuksel:2013, quevedo:2014,ostergaard:2016,peters:2016} also took into account achievable data rates due to communications. 
In \cite{sinopoli:2004},  it was shown that for a given unstable LTI system, there exists a critical limit on the packet dropout rate beyond which the system cannot be stabilized in the usual mean-square sense. 
To go beyond this critical limit, several techniques have been proposed ranging from error correction codes~\cite{singleton:1964,schulman:1996,sahai:2006,ostrovsky:2009,khina:2016} and multiple descriptions~\cite{ostergaard:2016}  to packetized predictive control~\cite{quevedo:2007} to name a few.

Error-correction codes are  often designed from a purely open-loop communications perspective, where the erasure probability of the channel is taken into account in the design but the plant properties are ignored.
For example, conventional $(n,k)$ erasure channel codes, take $k$ \emph{source} packets and outputs $n$ \emph{channel} packets. If any $k$ of the channel packets are received, the original $k$ source packets can be completely recovered. If more than $k$ packets are received, the additional received data packets are not useful since they do not contain any further information about the plant state than what is already known. Finally, if less than $k$ packets are received, the source packets can generally not be recovered and all the transmitted information is in this case wasted. While these erasure codes can be shown to be capacity achieving in the open loop case, they are not always sufficient for guaranteeing stability in the closed-loop case. In particular, 
it is well established that Shannon's information theoretic notion of channel capacity is generally not sufficient to fully characterize the rate required for stability. Instead, \cite{sahai:2006} introduced the concept of anytime capacity to characterize the channel rate required for stability. In principle, anytime capacity describes the maximal rate of a causal error correction code having a probability of error that decays at least  exponentially with the delay \cite{sahai:2006}. Practical anytime reliable codes have been presented in e.g.,~\cite{sukhavasi:2016,khina:2016}.

An alternative to error correction codes is multiple descriptions \cite{gamal:1982}, which combine source and channel coding. With multiple descriptions, the source is encoded into a number of descriptions, which are individually transmitted over the channel. There is no priority on the descriptions, and any subset of the descriptions can be jointly decoded to achieve a desired performance. 
Multiple descriptions were, for example, used for state-estimation in \cite{jin:2006} and combined with packetized predictive control in~\cite{ostergaard:2016}. One of the problems with multiple descriptions is that it is generally very hard to design good multiple--description codes. Another problem is that the descriptions generally contains redundant information except in the limit of vanishing data rates or when used in the extreme asymmetric situation, where the descriptions are prioritized and a successive refinement scheme is obtained. If one is able to construct a successive refinement source coder, then it was shown in \cite{yeung:1996} and \cite{puri:1999}, that the layers in the successive refinement code can be combined with traditional error correction codes in order to obtain a (sub optimal) multiple--description code. It was recently shown that a combination of successive refinement and multiple descriptions with feedback becomes   rate-distortion optimal under certain asymptotics \cite{ostergaard:2021}.  

In this paper, we propose a new type of practical delayless erasure codes for control over digital erasure channels, which takes the loss probability of the channel as well as the stability constraints of the system into account in the design. 
The current work extends our conference contribution presented in \cite{ostergaard:2021b} by including detailed descriptions of the construction of the stabilizing erasure codes, stability results for time-invariant and time-varying coding policies, and extensive simulations. 

We focus on discrete-time LTI plants, stationary Gaussian disturbances, Gaussian initial state, scalar-valued control inputs and sensor outputs. Thus, the plant state can have an arbitrary dimensionality but the control signal as well as the output of the plant are both scalar valued. The encoder is not aware of the erasures. We use variable-rate stochastic and deterministic coders and we are interested in obtaining stability and a desired output variance hereafter referred to as control performance. Towards that end, we provide lower bounds on the required coding rates, and design practical schemes having average bitrates that are within a constant gap from the bounds. We demonstrate that stabilizing erasure codes can be obtained from properly designed independent encodings~\cite{ostergaard:2021} or multiple descriptions~\cite{gamal:1982}. Specfically, for a given LTI plant we design a $(k,k')$ stabilizing code such that when combining any $k'$ descriptions of the code, the resulting $\mathrm{SNR}$ is above a critical limit, which guarantees that the decoded control signal contains sufficient information to stabilize the plant. If you receive more than $k'$ descriptions, the $\mathrm{SNR}$ is further improved leading to a better control performance. Finally, if you receive less than $k'$ packets, it is still possible to decode the source, however, the resulting $\mathrm{SNR}$ is less than that required for stability. By use of Markov jump linear system theory~\cite{costa:2005}, we also show how to design the $(k,k')$ code so that stability is achieved if on average at least $k'$ packets are received. This is useful for networks with random erasures.

We show that  simple codes based on independent encodings, are asymptotically efficient for stabilizing "nearly" stable plants. In general, for unstable plants, it is advantageous to use a design based on multiple descriptions. In a simulation study, we demonstrate that for the same sum-rate and delay, it is possible to achieve a significant gain in control performance over that which is possible with repetition coding. 

The paper is organized as follows. Section II provides the necessary background information on networked control systems and dithered quantization that will be needed in Section IV. Section III introduces the coding policies. Section IV contains the main contributions which are the theoretical and practical foundation for constructing stabilizing erasures codes and assessing their stability and efficiency. Section V contains a simulation study and the conclusions are in Section VI. All proofs are deferred to the appendix.

%
%
%
%
%
%
%
%
%
%
%
%
%
%
%
%
%

\section{Background}

Let us begin by  considering the  networked control system presented in \cite{silva:2016}, and which is shown in Fig.~\ref{fig:system} (left). Here $P$ is an LTI plant that is open-loop unstable, $u$ is the scalar control input, and $y$ is the scalar plant sensor output. The external disturbance is denoted by $d$ and $e$ is an error signal that is related to the control performance. The plant output $y$ is to be encoded by the causal encoder $\mathcal{E}$, transmitted over the ideal noise-less digital channel, and then decoded by the causal decoder $\mathcal{D}$. The encoder--decoder pair $(\mathcal{E},\mathcal{D})$ also contains the controller. Thus, the output of the decoder is the control signal to  the plant. 

For a fixed data rate of the coder, the control performance is measured by the output variance $\sigma_e^2$ of $e$. We have the following linear input-output relationship through the plant $P$:
\begin{equation}\label{eq:io}
\begin{bmatrix}
e \\
y
\end{bmatrix} 
= 
\begin{bmatrix}
P_{11} & P_{12} \\
P_{21} & P_{22} 
\end{bmatrix}
\begin{bmatrix}
d \\ u
\end{bmatrix}.
\end{equation}

It was shown in \cite{silva:2016}, that if the initial state $x_0$ and the external disturbances are arbitrarily colored but jointly Gaussian, then a theoretically optimal encoder--decoder pair constitute a linear system + noise. This implies that the system in Fig.~\ref{fig:system} (left) can be modelled by the \emph{linear} system shown in Fig.~\ref{fig:system} (right). In this system, $F$ and $L$ are LTI systems, and $q$ is an additive white Gaussian noise, which simulates the coding noise due to source coding. In practice, scalar quantizers do generally not result in Gaussian noise. This deviation from Gausianity implies that the model is not exact for practical schemes, which influences  the resulting coding rate as explained in Section II.A.
Using the model of Fig.~\ref{fig:system}(right) yields the following relationship \cite{silva:2016}:
\begin{equation}\label{eq:system}
u = Fw, \quad w= v+q, \quad v = L  
\begin{bmatrix}
z^{-1}w \\
y
\end{bmatrix}, L = [L_w, L_y],
\end{equation}
where $z^{-1}$ indicates a one-step delay operator.
The signal-to-noise ratio $\gamma$ of the system is defined as:
\begin{equation}
\gamma \triangleq \frac{\sigma_v^2}{\sigma_q^2}.
\end{equation}

The system $P$ in Fig.~\ref{fig:system} (right) is described by the following state-space recursions:
\begin{align}\label{eq:state_recursion}
x(i+1) &= A x(i) + B_1 d(i)  + B_2 u(i) \\
e(i) &= C_1 x(i) + D_{11} d(i) + D_{12} u(i) \\ \label{eq:yi}
y(i) &= C_2 x(i) + D_{21} d(i),
\end{align}
where the matrices $A, B_1, B_2, C_1, C_2, D_{11}, D_{12}, D_{21}$ are time invariant and of appropriate dimensions.

It was shown in \cite{silva:2016}, that for any proper LTI filters $F$ and $L$ that make the system in Fig.~\ref{fig:system} (right) internally stable and well-posed, we have the following explicit expressions:
\begin{align} \label{eq:gamma}
\gamma &= \|S - 1\|^2 + \frac{1}{\sigma_q^2}\| L_y P_{21} S\|^2 \\ \label{eq:perf}
\sigma_e^2 &= \|P_{11} + P_{12}K(1-P_{22}K)^{-1}P_{21}\|^2 + \|P_{12}FS\|^2 \sigma_q^2 \\ \label{eq:S}
S &= (1-L_w z^{-1} - P_{22} F L_y)^{-1}\\
K &= F L_y(1 - L_w z^{-1})^{-1}.
\end{align}

To find the optimal filters $(F,L)$ that minimize $\sigma_e^2$ subject to a constraint on $\gamma$, one needs to solve a convex optimization problem \cite{silva:2016}. A lower bound on the minimal average data rate $\mathcal{R}$ achievable when using optimal filters $(F,L)$ is \cite{silva:2016}:
\begin{equation}\label{eq:min_rate1}
\mathcal{R} \geq \frac{1}{2}\log_2( 1 + \gamma).
\end{equation}
It is clear from \eqref{eq:gamma} that asymptotically as $\sigma_q^2\to \infty$, $\gamma \to \|S-1\|^2$, which shows that the minimum $\mathrm{SNR}$ required for stability is:
\begin{equation}\label{eq:min_snr}
\gamma > \|S-1\|^2,
\end{equation}
and the minimum rate required for stability is:
\begin{equation}\label{eq:min_rate}
\mathcal{R} > \frac{1}{2}\log_2(1 + \|S-1\|^2).
\end{equation}

\subsection{Entropy constrained  dithered quantization}
In Fig.~\ref{fig:system} (right), the signal $q$ models the quantization noise due to coding. In particular, it is assumed that $w=v+q$, where $q$ is independent of $v$. Moreover, it is assumed that $w$ has a density. In practice, the output of a quantizer must be discrete to be conveyed over the digital channel. However, for a stochastic quantizer, the reconstruced output at the decoder can easily be made continuous. In this work, we assume the existence of a subtractively dithered and entropy constrained scalar quantizer $\mathcal{Q}_\Delta : \mathbb{R}\times \mathbb{R} \to \Delta\mathbb{Z}$  having step-size $\Delta \in \mathbb{R}_+$ \cite{zamir:2014}. This is a stochastic quantizer, which relies upon a pseudo-random dither signal $z\in \mathbb{R}$, which is uniformly distributed, $z \sim \mathcal{U}(-\Delta/2, \Delta/2)$, and known at both the encoder and decoder. Specifically, let $v\in \mathbb{R}$,  then:
\begin{align}
v_c &=  \mathcal{Q}_\Delta(v + z)  =  \bigg\lceil  \frac{  (v + z)}{\Delta} \bigg\rfloor \Delta  \in \Delta \mathbb{Z},
\end{align}
where $\lceil \cdot \rfloor$ denotes rounding to nearest integer, and where $z$ is independent of $v$. 
Since the quantizer is unbounded, we avoid the difficulty of having to deal with overload distortion, which could potentially lead to instability. 
At the decoder, the reconstructed signal is given by:
\begin{align}
w &= v_c - z, 
\end{align}
where $w\in \mathbb{R}$ has a well-defined density. Since $z$ is independent of $v$,  the error $z' \triangleq v-w = v- v_c +z$ is also independent of $v$ and is distributed similar to $z$ \cite{zamir:2014}. At any data rate, the additive model $w = v + q$ of Fig.~\ref{fig:system} (right) is valid by choosing  $q = -z' = v_c - v - z$.The conditional output entropy of the quantizer given the dither is equal to the  mutual information between the input and the output of the quantizer \cite{zamir:2014}:
\begin{align*}
H(v_c | z) &= I(v ; w) = I( v; v+q) \leq   I( \bar{v}; \bar{v}+\bar{q}) + D(q\| \bar{q} )  \\
&= 0.5\log_2\bigg( 1 + \frac{\sigma_v^2}{\sigma_q^2}\bigg) + \frac{1}{2}\log_2(\pi e/6),
\end{align*}
where $D(\cdot\| \cdot)$ denotes the information divergence~\cite{cover:2006}, and the bar-notation refers to the Gaussian counterparts of the variables, i.e., $\bar{v}$ denote a Gaussian random variable, which has the same $1^{\mathrm{st}}$ and $2^{\mathrm{nd}}$ moments as that of $v$ \cite[Lemma 2]{derpich:2008}. If $q$ is uniformly distributed, then $D(q\| \bar{q} ) =\frac{1}{2}\log_2(\pi e/6)$.

The discrete output $v_c$ of the quantizer is mapped to a bitstring using an entropy coder \cite{shannon:1948}. An entropy coder is a fixed-to-variable length mapping, where high probability symbols are encoded with a short bitstring, and low probability symbols are mapped to longer bitstrings \cite{shannon:1948}. The average operational data rate $R$ of a uniquely decodable scalar entropy code is bounded by: $H(v_c|z) \leq R \leq H(v_c|z)  + 1$ \cite{shannon:1948}. Thus, the average data rates of the system in Fig.~\ref{fig:system} (right) when using a scalar quantizer followed by scalar (memoryless) entropy coding is bounded by \cite{silva:2016}: 
\begin{align}\notag
\frac{1}{2}\!\log_2\!\!\bigg( 1 + \frac{\sigma_v^2}{\sigma_q^2}\bigg)  \leq R  \label{eq:min_rate2}
< \frac{1}{2}\!\log_2\!\!\bigg( 1 + \frac{\sigma_v^2}{\sigma_q^2}\bigg) +  \frac{1}{2}\!\log_2\!\!\bigg(\frac{\pi e}{6}\bigg) + 1.
\end{align}

\section{Causal Encoders and Decoders}\label{sec:decoder}
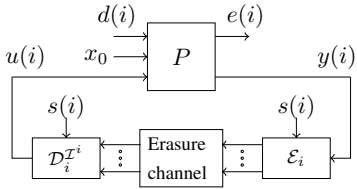
\begin{figure}
\centering
\scalebox{0.9}{
\begin{tikzpicture}

\draw   
   (2,3) rectangle (3,2)
	node[pos=0.5]{$P$};

\draw [->] (1.5,2.8) -- (2,2.8)
	node[pos=0.1,above]{$d(i)$};
\draw [->] (1.5,2.5) -- (2,2.5)
	node[pos=0.1,left]{$x_0$};
\draw [->] (0,2.2) -- (2,2.2)
	node[pos=0.1,above]{$u(i)$};
\draw [-] (3,2.2) -- (5,2.2)
	node[pos=0.9,above]{$y(i)$};
\draw [->] (3,2.8) -- (3.5,2.8)
	node[pos=0.9,above]{$e(i)$};

\draw [->] (5,2.2) -- (5,1) -- (4.7,1);
\draw [-] (0,2.2) -- (0,1) -- (0.3,1);

\draw [<-] (3.1,0.8) -- (3.7,0.8);
\draw (3.4,1.1) circle (0.5pt); 
\draw (3.4,1) circle (0.5pt); 
\draw (3.4,0.9) circle (0.5pt); 
\draw [<-] (3.1,1.2) -- (3.7,1.2);

\draw   
   (0.3,0.7) rectangle (1.3,1.3)
	node[pos=0.5]{\footnotesize $\mathcal{D}^{\mathcal{I}^i}_i$}; 

\draw   
   (3.7,0.7) rectangle (4.7,1.3)
	node[pos=0.5]{\footnotesize $\mathcal{E}_i$};

\draw [<-] (1.3,0.8) -- (1.9,0.8) ;
\draw (1.6,1.1) circle (0.5pt); 
\draw (1.6,1) circle (0.5pt); 
\draw (1.6,0.9) circle (0.5pt); 
\draw [<-] (1.3,1.2) -- (1.9,1.2);

\draw [<-] (0.8,1.3) -- (0.8,1.6)
	node[pos=0.5,above]{$s(i)$};

\draw [<-] (4.2,1.3) -- (4.2,1.6)
	node[pos=0.5,above]{$s(i)$};



\node[draw, text width=9.9mm] at (2.5,1) {\footnotesize Erasure channel};


\end{tikzpicture}
}

\caption{Noisy LTI system $P$ that is controlled over a digital erasure channel. The causal encoder $\mathcal{E}_i$ produces $k$ outputs for each input symbol $y(i)$. The causal decoder $\mathcal{D}^{\mathcal{I}^i}_i$ receives $k' \leq k$ outputs and produces a single control signal $u(i)$. }
\label{fig:md_system}
\vspace{-5mm}
\end{figure}

A sketch of the proposed system can be seen in Fig.~\ref{fig:md_system}. 
The encoder $\mathcal{E}_i : \mathbb{R}^i  \times \mathcal{S}^i \to \mathcal{Y}_c^k$ at time $i$  is a (possibly) time-varying causal one-to-many map, which at each time instance produces $k$ outputs, that is:
\begin{align}
(y_c^{(1)}(i), \dotsc, y_c^{(k)}(i))  &= \mathcal{E}_i(y^i, s^i),
\end{align}
where $y^{(j)}_c(i) \in \mathcal{Y}_c$ denotes the $j$th output of the encoder at time $i$, and $y^i = y_1,\dotsc, y_i$ indicates that the encoder is only using the sequence of current and past plant outputs. 
The sequence $s^i$ denotes side information. Thus, the encoder can be randomized via the side information, which for example allows one to obtain a stochastic encoder. We make use of this property, where in particular we  let $s^i$ be a dither signal, which is known both at the encoder and decoder. In this case the we let $\mathcal{S} = \mathbb{R}$. 

The expected length (in bits) of $y^{(j)}_c(i)$ is denoted $R^{(j)}(i)$ and the average data rate $\mathcal{R}^{(j)}$ of the $j$th description is:
\begin{align}
\mathcal{R}^{(j)} = \lim_{n\to \infty} \frac{1}{n} \sum_{i=1}^n R^{(j)}(i).
\end{align}
The time-averaged sum-rate is:
\begin{align}
\mathcal{R}_s = \sum_{j=1}^k  \mathcal{R}^{(j)}.
\end{align}

For $k=1$, we obtain a single--description system. We reserve the notation $\mathcal{R}$ to denote the average data-rate obtained for a single--description system:
\begin{align}
\mathcal{R} = \lim_{n\to \infty} \frac{1}{n} \sum_{i=1}^n R(i).
\end{align}
If $k=1$, then $\mathcal{R} = \mathcal{R}_s$.

Let $\mathcal{I}(i) \subseteq \{1,\dotsc, k\}$ denote the set of indices of the received descriptions at time $i$. Recall that at each time instance, $k$ descriptions are produced and transmitted over the digital erasure channels. As an example, let $k=3$ and assume that the second description, $y_c^{(2)}(i)$, is lost at time $i$. Then $\mathcal{I}(i) = \{1,3\}$. Let $\mathcal{I}^i = \mathcal{I}(1), \mathcal{I}(2), \dotsc, \mathcal{I}(i)$. We are now in a position to define the set of causal decoders at time $i$, i.e., $\mathcal{D}^{\mathcal{I}^i}_i :  \mathcal{Y}_c^{\mathcal{I}^i}  \times \mathcal{S}^{\mathcal{I}^i} \to \mathbb{R},\ \forall \mathcal{I}^i \subseteq \{1,\dotsc, k\}^i$. For a particular choice of decoder, say $\mathcal{D}^{\mathcal{I}^i}_i$, the reconstructed signal $u(i) \in \mathbb{R}$ at time $i$ is given by:
\begin{align}
u(i) = \mathcal{D}^{\mathcal{I}^i}_i (  y_c^{\mathcal{I}^i} , s^{\mathcal{I}^i}). 
\end{align}

\section{Stabilizing Error Correction Codes}

We first introduce some definitions, which are needed in the sequel. 
When we refer to stability we are concerned with mean-square stability (MSS) defined as $\sup_n \mathbb{E}[ x(n)x(n)^T] < \infty$. The term control performance refers to a bounded steady state variance $\sigma_e^2$ \eqref{eq:perf}, which implicitly assumes that the system becomes asymptotically wide sense stationary (AWSS) (which further implies MSS).  A random process is said to be AWSS if and only if there exists $\mu_x$ and $C_x(\tau)$ such that  $\lim_{n\to \infty} \mathbb{E}[ x(n)] = \mu_x$ and  
$\lim_{n\to \infty} \mathbb{E}[ (x(n+\tau) - \mathbb{E}[x(n+\tau)]) (x(n+\tau) - \mathbb{E}[x(n+\tau)])^T] = C_x(\tau), \forall \tau$, independently of $x(0)$ \cite{costa:2005}.

\begin{definition}
We  denote by $(F,L,P,\gamma)$ a linear system on the form shown in Fig.~\ref{fig:system} (right) that has average single-rate $\mathcal{R}=0.5\log_2(1+\gamma)$ and control performance $\sigma_e^2$, where $\sigma_e^2$ is given by \eqref{eq:perf}.
\hfill$\triangle$
\end{definition}

\begin{definition}
 A $(k,k')$ stabilizing code for the system $(F,L,P, \gamma)$ produces $k$ descriptions such that using any $k'$ of them is sufficient to stabilize the system.
\hfill$\triangle$
\end{definition}



To quantify the efficiency of a $(k,k')$ stabilizing code when used on a particular system $(F,L,P,\gamma)$, we compare the average sum-rate $\mathcal{R}_s$ of the $k$ descriptions to the average single-description rate $\mathcal{R}$ required for a single-description code to achieve the same control performance as that obtained when using all $k$ descriptions (without erasures). In the linear Gaussian case, the efficiency can be assessed by simple means as shown in the definition below.

\begin{definition}
The efficiency $\eta$ of a $(k,k')$ stabilizing code for the system $(F,L,P,\gamma)$ is defined as:
\begin{equation}\label{eq:eta}
\eta \triangleq  
\frac{ \log_2( 1 + \tilde{\gamma})}{k\log_2( 1  + \hat{\gamma})},\quad 0\leq \eta \leq 1,
\end{equation}
where $\hat{\gamma}$ is the $\mathrm{SNR}$ when using any single description, and $\tilde{\gamma}$ is the $\mathrm{SNR}$ when combining all $k$ descriptions. \hfill$\triangle$
\end{definition}

When measuring efficiency in \eqref{eq:eta}, we need to make sure that we compare data rates of systems having similar control performance (in terms of $\sigma_e^2$). The best performance of a $(k,k')$ stabilizing code is obtained when using all $k$ descriptions, which results in an $\mathrm{SNR}$ of $\tilde{\gamma}$. The rate of each description is $0.5\log_2(1+\hat{\gamma})$ and the sum-rate is $0.5k\log_2(1+\hat{\gamma})$. On the other hand, when $k=1$  we need a data rate of $0.5\log_2(1+\tilde{\gamma})$ to achieve an $\mathrm{SNR}$ of $\tilde{\gamma}$. 

For a classical $(n,k)$ error correction code that produces $n$ outputs for each $k$ input blocks, the efficiency is $k/n$, and the delay is $k-1$. A repetition code that transmits the same source block $k$ times has efficiency $1/k$ and zero delay. The delayless stabilizing codes that we propose are able to improve upon the efficiency of repetition codes due to the property that descriptions can synergistically improve upon each other.

\subsection{Stabilizing codes based on independent encodings}\label{sec:indp}
In this section, we develop the theoretical foundation for stabilizing erasure codes based on independent encodings. The practical design follows in Section~\ref{sec:practical}.

\begin{definition} \label{def:encodings}
Let $y_i = v + q_i, i = 1,\dotsc, k$. If $y_j$ and $y_i, i\neq j,$ are conditionally independent given $v$, then we refer to $y_1,\dotsc, y_k$ as \emph{independent encodings}~\cite{ostergaard:2021}.
\end{definition}

\begin{lemma}\label{lem:indp}
\emph{Consider the system $(F,L,P,\gamma)$, which is illustrated in Fig.~\ref{fig:system} (right). Let $v$ be Gaussian and let $y_i = v + q_i, i = 1, \dotsc, k$, be $k$ independent encodings of $v$,  
where $q_i, i=1,\dotsc, k,$ are mutually independent,  zero-mean, and having a common finite variance $\sigma^2$. If for some $1\leq k' \leq k$, the common variance satisfies
\begin{equation}\label{eq:code1}
\sigma^2\leq  \frac{\gamma k'\|L_yP_{21}S\|^{2}}{ \|S-1\|^{2} (\gamma - \|S-1\|^2)},
\end{equation}
where $S$ is given in \eqref{eq:S}, then $y_i, i=1,\dotsc, k$, form a $(k,k')$ stabilizing code for the system $(F,L,P,\gamma)$. }
\end{lemma}

The following lemma provides a lower bound on the sum-rate required for a $(k,k')$ stabilizing code. We note that this bound is for the case where the individual descriptions are also helpful by themselves. If one is not interested in the control performance when receiving less than $k'$ descriptions, then the sum-rate can generally be further reduced by use of distributed source coding techniques such as Slepian-Wolf coding \cite{slepian:1973}. However, at low coding rates, the bound becomes asymptotically optimal as is shown by Lemma~\ref{lem:eta_indp}. Note that the lemmas in this subsection establish lower bounds. Achievability of the lower bounds are considered in Section~\ref{sec:practical}.

\begin{lemma}\label{lem:R}
\emph{The  average sum-rate $\mathcal{R}_s$ of a $(k,k')$ stabilizing code based on independent encodings for the system $(F,L,P,\gamma)$ is lower bounded as:}
\begin{equation}
 \mathcal{R}_s \geq \frac{k}{2}\log_2\bigg(1 + \frac{\|S+1\|^2}{k'}\bigg).
\end{equation}
\end{lemma}

\begin{lemma}\label{lem:eta_indp}
\emph{Consider the system $(F,L,P,\gamma)$. The efficiency of a minimum sum-rate $(k,k')$ stabilizing code based on independent encodings is given by:
\begin{equation}
\eta =  \frac{\log_2(1 + k(k')^{-1}\,\|S-1\|^{2}) }
{k\log_2(1 + (k')^{-1}\,\|S-1\|^{2})},
\end{equation}
and the code is asymptotically efficient in the sense of:}
\begin{equation}
\lim_{\|S-1\|^2 \to 0} \eta = 1.
\end{equation}
\end{lemma}

The second part of Lemma~\ref{lem:eta_indp} considers the situation where the plant is either stable or nearly stable, i.e., the unstable poles are near the unit circle. In this case, the coding rates are arbitrarily small, and the $k$ descriptions of the $(k,k')$ stabilizing code become mutually independent. Thus, there is no redundancy by using $k$ descriptions each of rate $R/k$ over a single description of rate $R$  \cite{ostergaard:2021}.

\begin{lemma}\label{lem:indp4}
\emph{Consider the system $(F,L,P,\gamma)$. The control performance (in terms of $\sigma_e^2$) for this system when using $\ell \geq k'$ descriptions of a minimum sum-rate $(k,k')$ stabilizing code based on independent encodings is:}
\begin{align}\notag
\sigma_e^2 &= \|P_{11} + P_{12}K(1-P_{22}K)^{-1}P_{21}\|^2 \\
&\quad +
 \frac{k'\, \gamma  \|P_{12}FS\|^2 \|L_yP_{21}S\|^{2}}{\ell\, \|S-1\|^{2} (\gamma - \|S-1\|^2) }
 			,\quad \ell=k', \dotsc, k.
\end{align}
\end{lemma}

%
%

%

\subsection{Stabilizing codes based on multiple descriptions}\label{sec:md}
In this section, we develop the theoretical foundation for stabilizing erasure codes based on multiple descriptions. The practical design follows in Section~\ref{sec:practical}.

It is possible to introduce negative correlation between the quantization noises $q_i, i=1,\dotsc, k,$ of the encodings in Definition~\ref{def:encodings}, which makes it possible to exploit the benefits of multiple descriptions. Of course, zero correlation is a special case of multiple descriptions, which is usually referred to as the no excess marginal rate case~\cite{Zhang1995MultipleDS}. 
When introducing correlation, the sum-rate is no longer simply just given by the sum of the optimal marginal (description) rates. The sum-rate also becomes a function of the amount of correlation introduced; the greater (negative) correlation, the greater  sum-rate \cite{ozarow:1980}. 

\begin{lemma}\label{lem:md}
\emph{Consider the system $(F,L,P,\gamma)$, which is illustrated in Fig.~\ref{fig:system} (right). Let $v$ be Gaussian and let $y_i = v + q_i, i = 1, \dotsc, k$, 
where $q_i, i=1,\dotsc, k,$ are zero-mean and of finite variance $\sigma^2$, and pairwise correlated with correlations coefficient $-\frac{1}{k-1}<\rho \leq 0$. 
If for some $k'$ and $\rho$, the common variance $\sigma^2$ satisfies
\begin{equation}\label{eq:code2}
\sigma^2\leq  \frac{\gamma k'\|L_yP_{21}S\|^{2}}{ \|S-1\|^{2} (\gamma - \|S-1\|^2)(1+(k'-1)\rho)},
\end{equation}
where $S$ is given in \eqref{eq:S}, then $y_i, i=1,\dotsc, k$, form a $(k,k')$ stabilizing code for the system $(F,L,P,\gamma)$. }
\end{lemma}

Let $\rho \in (\frac{-1}{k-1},0]$ be the common correlation coefficient between all noise pairs $q_i,q_j, \forall i\neq j$, and let $\sigma^2$ be their common variance. If we are only interested in the control performance when receiving $k'$ descriptions or all $k$ descriptions, then the sum-rate $\mathcal{R}_s$ can be explicitly lower bounded by  \cite{pradhan:2004}:
\begin{align}\notag
\mathcal{R}_s  &\geq \log_2\!\! \bigg( \frac{k' + \sigma^2(1+(k'-1)\rho)}{\sigma^2(1-\rho)} \bigg)^{\frac{k}{2k'}} 
  \bigg(\frac{1-\rho}{1+(k-1)\rho} \bigg)^{\frac{1}{2}},
\end{align}
where it is assumed the source is standard normal, and where equality is achieved if the quantization noises could be made Gaussian.

%
%
%
%
%
%
%
%
%
%

\subsection{Practical Design of Stabilizing Codes} \label{sec:practical}
In this section we show how to construct  stabilizing erasures codes based on the theory of independent encodings and multiple descriptions introduced in Sections~\ref{sec:indp} and \ref{sec:md}, respectively. 

\subsubsection{Independent encodings}
We use a subtractively dithered scalar quantizer, which is a stochastic quantizer that provides different outputs, when encoding the same source multiple times \cite{zamir:2014}. We use this to form the $k$ \emph{independent} encodings by encoding the output $v$ of Fig.~\ref{fig:system} (right) using a  subtractively dithered scalar quantizer $\mathcal{Q}_\Delta$ with step-size $\Delta$: 
\begin{equation}
w_i = \mathcal{Q}_\Delta(v + \xi_i)-\xi_i, i = 1,\dotsc, k,
\end{equation}
where $\xi$ denotes the dither signal. We choose  $\Delta$ apropriately to ensure that the resulting $\mathrm{SNR}$, when receiving $k'$ descriptions is sufficient for stabilization. 
The quantized sensor output is  entropy coded using a scalar (memoryless) entropy coder. 
The variance $\sigma^2$ of the quantization error obtained when reconstructing using only one description is given by $\sigma^2 = \Delta^2/12$. The reduction in noise variance is linear in the number of descriptions $\ell$ and given by $\sigma^2(\ell) = \sigma^2/\ell$.
The independent encodings do not result in data rates that reaches the lower bound of Lemma \ref{lem:R}. Indeed, they suffer from a rate loss: 

i) The scalar quantizer produces uniformly distributed noise, and the lower bound on the data rates assumes Gaussian distributed noise. This loss is at most $0.25$ bits/dim. \cite{zamir:2014}. 

ii) 
It is well known that a scalar memoryless entropy coder suffers a rate loss of at most 1 bit per sample \cite{shannon:1948}.

iii) A static entropy coder is designed for a specific distribution. However, due to erasures, the distribution of the quantized sensor output  could be time varying. We design the entropy coder by assuming i.i.d.\ packet losses, and AWSS distributions. The rate loss observed in our simulations is fairly small.  

iv) 
For each quantized sensor output, the entropy coder needs to condition its output upon the particular realization of the dither signal.  In practice, one can choose to discretize the number of possible dither values to a limited set or e.g., simply use an unconditional entropy coder that ignores the dither values. In the simulation section, we choose the latter option. 

\subsubsection{Multiple descriptions}

It is not straight-forward to obtain correlated noises between the descriptions, and we use here the approach described in~\cite{ostergaard:2006}, which is based on nested quantizers and index assignments.  The source is first quantized using a fine-grained scalar \emph{deterministic} uniform quantizer $\mathcal{Q}_\Delta$ with step size $\Delta$. We refer to $\mathcal{Q}_\Delta$ as the central quantizer. 
Then, a one-to-many \emph{index-assignment} map $\phi: \Delta\mathbb{Z} \to \Delta_s\mathbb{Z}^k$ is applied, which maps the quantized value $b \in \Delta \mathbb{Z}$ to a $k$-tuple $(a_1,\dotsc, a_k) \in \Delta_s \mathbb{Z}^k$, where $\Delta_s/\Delta \in \mathbb{N}$ denotes the nesting ratio. 
Specifically, let $v\in \mathbb{R}$ then:
\begin{align}
b &= \mathcal{Q}_\Delta(v)  \in \Delta \mathbb{Z}\\
(a_1, \dotsc, a_k) &= \phi( b ) \in \Delta_s \mathbb{Z} \times \cdots \times \Delta_s\mathbb{Z}.
\end{align}
The $k$ quantized values $a_j, j=1,\dotsc, k,$ are individually entropy encoded and transmitted separately over the network. 

Let $\mathcal{I}(i)$ denote the indices of the received descriptions at time $i$ and let $J_i=|\mathcal{I}(i)|$ be the number of received descriptions. To simplify the notation, we (for the moment being) ignore the dependency on $i$ and consider a single time instant. From the recevied descriptions, we can recover (by seperate  entropy decoding) $\bar{a} = (a_{j_1,}, \dotsc, a_{j_J})$, where $j_1,\dotsc, j_{J} \in \mathcal{I}$. If $J = k$, then $\bar{a} =  (a_1,\dotsc, a_k)$ otherwise $\bar{a} \subset (a_1,\dotsc, a_k)$. If $J=0$ then $\bar{a} = \emptyset$.

Given all $k$ points in a $k$-tuple $(a_1,\dotsc, a_k)$, the map $\phi$ is invertible and the point $b$ of the central quantizer can be obtained, that is \cite{ostergaard:2006}:
\begin{align}
b = \phi^{-1}(a_1,\dotsc, a_k).
\end{align}
If $J<k$, then less than $k$ points of $(a_1,\dotsc, a_k)$ are available. In this case, the reconstruction is given by the average of the received points \cite{ostergaard:2006}. The output $w$ is then obtained using the following decoding policy:
\begin{align}
w &= \begin{cases}
b, &  J=k, \\
\frac{1}{J}\sum_{j\in \mathcal{I}} a_j , & k>J>0, \\
\mathbb{E}[v], & \text{otherwise}.
\end{cases}
\end{align}

The design of $\phi$ can be done by solving a simple bipartite graph theoretical problem \cite{ostergaard:2006}. 
Specifically, let $r=\Delta_s/\Delta$ be an odd non-negative integer, let $\Lambda = \Delta \mathbb{Z}$, and let $\Lambda_s = \Delta_s \mathbb{Z}$. Finally, let $\Lambda_k =  r^2 \mathbb{Z}$. 
Then $\Lambda_k \subseteq \Lambda_s \subseteq \Lambda$.  Define the discrete set $\mathcal{V}$ as follows:
\begin{align}
\mathcal{V} = \{ a \in \Lambda |  |a| \leq | a-a' |, \forall a' \in \Lambda_k \}, 
\end{align}
which implies that $\mathcal{V}$ contains the $r^2$ points of $\Lambda$ which are closer to the origin than to any other point in $\Lambda_k$. It may be deduced that  $|\mathcal{V} \cap \Lambda_s| = r$.

Let $(a_1, \dotsc, a_k) \in \Lambda^k$ be a $k$-tuple that contains $k$ points that each belong to $\Lambda_s$, i.e., $a_i \in \Lambda_s, i=1,\dotsc, k$. Let $\mathcal{S}(a_1)$ be the set of all distinct $k$-tuples in $\Lambda_s^k$ whose first element is $a_1$ and where the pairwise distances between the elements are no greater than $\varphi_{r,k} >0$, i.e.,
\begin{align}
\mathcal{S}(a_1) = \{ (a_1,\dotsc, a_k) \in \Lambda_s^k | \, \| a_i - a_j \| \leq \varphi_{r,k},\forall i,j \},
\end{align}
where $\varphi_{r,k}$ depends upon $r$ and $k$ and is chosen sufficiently large to ensure that $|\mathcal{S}(a_1)| \geq r$. We use $\mathcal{S}(a_1)$ to construct the super set $\mathcal{S}$:
\begin{align}
\mathcal{S} = \{ S(a_1) | a_1 \in \mathcal{V} \cap \Lambda_s \},
\end{align}
which contains at least $r^2$ distinct $k$-tuples that all have their first coordinate within $\mathcal{V}$. 

We are now in a position to construct the non-linear map $\phi$. Towards that end, we form a bi-partite linear assignment optimization problem \cite{west:2001}, where $r^2$ out of the $|\mathcal{S}|$ $k$-tuples in $\mathcal{S}$ are assigned to the $r^2$ points in $\mathcal{V}$. Let the cost $c(\bar{a}, \lambda)$ of a given $k$-tuple $\bar{a} \in \mathcal{S}$ when assigned to the point $\lambda \in \mathcal{V}$ be given by:
\begin{align}
c(\bar{a}, \lambda) = | \lambda -  k^{-1}\sum_{i=1}^k a_i|.
\end{align}
Then, an optimal assignment $\phi^*$ is one that minimizes:
\begin{align}
	\phi^* = \min_\phi  \sum_{\lambda \in \mathcal{V}}  c(\bar{a}=\phi(\lambda), \lambda),
\end{align}
where the minimization is over all possible  one-to-many maps $\phi: \mathcal{V} \to \mathcal{S}$ that satisfy $\phi^{-1}( \phi(\lambda)) = \lambda, \forall \lambda \in \mathcal{V}$.  

As an example consider the case where $\Delta = 1, \Delta_s = 3$, and $k=2$ or $k=3$. Then $\mathcal{V}$ contains the $9$ points illustrated in the columns entitled $b$ of Table \ref{tab:phi4} (left) and (right). Also shown in the tables are the assigned $9$ $k$-tuples for the case of $k=2$ (left) and $k=3$ (right). 
Another example is given in Table~\ref{tab:phi1} for the case of $\Delta=1, \Delta_s = 7$ and $k=3$, where $\mathcal{V}$ contains the $7^2=49$ points illustrated in the columns entiled $b$. The assigned $k$-tuples are also shown. Since all assigned $k$-tuples have their first coordinate in $\mathcal{V}$ they are so-called shift-invariant with respect to translations by $\lambda_k \in \Lambda_k$, i.e., $\phi(\lambda + \lambda_k) = \phi(\lambda) + \lambda_k, \forall \lambda_k \in \Lambda_k$. 

We now assess the rate and control performance of the multiple--description construction. 

Since the quantizer is a deterministic quantizer, it is not possible to analytically assess the data rate (entropy) and the variance of the quantization noise. An exception is under the condition of high resolutions, which in this case means in the limit where $\Delta, \Delta_s \to 0$ and $\Delta_s/\Delta \to \infty$. In this case the variance $\sigma^2(\ell)$ of the quantization noise of any single description ($\ell=1$)  is asymptotically given by \cite{ostergaard:2006}:
\begin{equation}\label{eq:rate_asymp}
\lim_{\Delta,\Delta_s\to 0} \lim_{\Delta_s/\Delta \to \infty} \Delta^{-2} \sigma^2(1) = \frac{1}{12},
\end{equation}
and the variance due to averaging $k>\ell>0$ descriptions is: 
\begin{align}\notag
&\lim_{\Delta,\Delta_s\to 0} \lim_{\Delta_s/\Delta \to \infty} \!\! \bigg(\!\frac{\Delta_s}{\Delta}\!\bigg)^{\!\!-2k/(k-1)} \frac{\Delta^{-2}}{12} \sigma^2(\ell) 
=  \frac{(k-\ell)}{2k\ell} \psi^2(k),
\end{align}
where $1\leq \psi(k)<2$ is an expansion factor that is related to $\varphi_{k,r}$. For $k=2, \psi(2) = 1$, and for $k=3, \psi(3) = 1.1547$ \cite{ostergaard:2006}. 
The sum-rate (entropy) $\mathcal{R}_s$ of all $k$ descriptions satisfies \cite{ostergaard:2006}:
\begin{align*}
\lim_{\Delta,\Delta_s\to 0} \lim_{\Delta_s/\Delta \to \infty} \mathcal{R}_s + k\log_2(\Delta_s) = k h(v),
\end{align*}
where $h(v)$ is the differential entropy of the input $v$ to the quantizer. 

At general resolutions in the non asymptotical case, a good approximation of the variance $\sigma^2(\ell)$ of the quantization error for $k\geq \ell > 0$ is given by \cite{ostergaard:2006}:
\begin{align}\label{eq:sigmaell}
\sigma^2(\ell) &\approx   \sigma^2 + \frac{(k-\ell)}{2k\ell} \frac{\Delta^2}{12}   \bigg(\frac{\Delta_s}{\Delta}\bigg)^{2k/(k-1)} \psi(k)^2,
\end{align}
where $\sigma^2 = \Delta^2/12$ is also the distortion obtained for $\ell = k$. Moreover, the sum-rate can be approximated by:
\begin{align}\label{eq:Rs_mdc}
 \mathcal{R}_s &\approx  k h(x) - k\log_2(\Delta_s)  =  \frac{k}{2}\log_2(2\pi e \sigma_v^2) - k\log_2(\Delta_s),
\end{align}
where the last equality follows by inserting the expression for differential entropy of a Gaussian source. At moderate to high data rates, the loss of the memoryless entropy coder is negligible, and we can expect operational sum-rates close to \eqref{eq:Rs_mdc}. We demonstrate this in the simulation section.

\begin{definition} \label{def:symmetric}
We use the term \emph{symmetric} code construction for the case where the where the rate is the same for all the descriptions, and 
the distortion does not depend upon which descriptions that are received but only upon the number of received descriptions.
\end{definition}

\textbf{Remark:} The code construction based on independent encodings is clearly symmetric. Similarly, 
from  \eqref{eq:sigmaell}  and \eqref{eq:Rs_mdc} it can be observed that the code construction based on multiple descriptions is asymptotically symmetric. 

\subsubsection{Practical efficiency}
Even a single-description system using a scalar quantizer suffers a rate loss. Thus, let us consider a notion of \emph{practical} efficiency where the sum-rate of the practical multiple description stabilizing code is compared to that of a single-description system using a scalar quantizer. 
\begin{lemma}\label{lem:prac}
\emph{The practical efficiency of the index-assignment based multiple descriptions stabilizing codes, asymptotically degenerate to that of the repetition code:}

\begin{equation}
 \lim_{\Delta \to 0} \frac{\mathcal{R}}{\mathcal{R}_s} = \frac{1}{k}.
 \end{equation}
\end{lemma}
\vspace{1mm}

In the non-asymptotical regime, where the data rates are finite, our simulations in Section~\ref{sec:sim} show that the efficiency of stabilizing codes based on multiple descriptions is significantly better than that of repetition codes.

\subsubsection{Designing stabilizing codes}

To design a stabilizing erasure code, one could consider the following two cases:

 1)  Assume that the bandwidth of the channel is sufficiently high to guarantee that at least $k'$ packets are successfully received. Then one could design the $(k,k')$ code without taking into account erasure probabilities as we do next.
 Assume we are given a plant $P$. Then we can find the minimal rate for stability from \eqref{eq:min_rate1} and \eqref{eq:min_rate2}. Let $\gamma$ be the minimum $\mathrm{SNR}$ required to achieve stability. Then it follows that we require $\sigma_q^2 < \sigma_v^2\gamma^{-1}$. Thus, to achieve stability when receiving $k'$ out of $k$ descriptions we can use \eqref{eq:sigmaell} and choose $(\Delta, \Delta_s)$ so that $\sigma^2(k') < \sigma_v^2\gamma^{-1}$. The resulting coding rate is then given by \eqref{eq:Rs_mdc}. In the simulation section, we provide a design example for a specific system $P$. 
 
2) If there is no guarantee on the minimum number of received descriptions, then one could choose $k,k', \Delta, \Delta_s$ so that the average variance of the quantization noise would be sufficiently small so as to stabilize the system. 
Towards that end, we refer to Lemma~\ref{lem:mjls} and Corollary~\ref{col:avgstb} below.

\subsection{Stability Analysis}
Let $\mathcal{I}$ be defined as in Section~\ref{sec:decoder}. If $\mathcal{I}(i) = \mathcal{I}(1), \forall i$, and if this is known to the encoder and decoder, then we can use a time-invariant coding policy, which uses a fixed encoder and decoder. In this case, we can rely on the results of \cite{silva:2016} in order to guarantee stability. Specifically, the encoder and decoder simply need to be chosen such that the combination of $\mathcal{I}(1)$ descriptions yields the desired $\mathrm{SNR}$. This can be done, for example, either by using independent encodings  or multiple descriptions for the design of the stabilizing codes. 

Let us now consider the case where $\mathcal{I}(i) \neq \mathcal{I}(1)$ for some $i\in \mathbb{N}$. Then we need to use a time-varying coding policy, where the choice of decoder depends upon the erasure patterns. A similar situation was considered in \cite{ostergaard:2016}, where it was shown that the system of \cite{ostergaard:2016} could be formulated as a Markov jump linear system (MJLS) \cite{costa:2005}. Moreover, 
it was shown in \cite{ostergaard:2016}, that if the sequence of erasure patterns was ergodic and independent of the external disturbances, then under certain conditions upon the sequence of multiple--descriptions encoder--decoder pairs, the system became AWSS. Whilst our setup here is different from that of~\cite{ostergaard:2016}, we can also here employ MJLS theory to prove stability, as is shown in the next lemma.

\begin{lemma}\label{lem:mjls}
Fix the system parameters $(F,L,P,\gamma)$, and a time-invariant symmetric encoder $\mathcal{E}_i(\cdot, \cdot)$. Moreover, assume i.i.d.\ packet losses, and let the choice of decoder at time $i$ only depend upon the erasure pattern at time $i$ via a jump variable $\xi_i$. Then one can form an MJLS:
\begin{align} \label{eq:mjls}
\bar{x}(i+1) 
&=\mathcal{A}(\xi_i)
\bar{x}(i)
+
 \mathcal{B}(\xi_i)
\begin{bmatrix}
d(i) \\
\bar{q}(i) 
\end{bmatrix},
\end{align}
which is AWSS whenever $\rho(\mathbb{A})<1$, and where $\bar{q}, \bar{x}, \mathcal{A}, \mathcal{B}$, and $\mathbb{A}$ are given in \eqref{eq:barq}, \eqref{eq:barx}, \eqref{eq:A}, \eqref{eq:B}, and \eqref{eq:AA}, respectively.
\end{lemma}

\begin{corollary}\label{col:avgstb}
Consider the system $(F,L,P,\gamma)$, and assume i.i.d.\ packet dropouts. 
Let  $p_s(\ell)$ be the probability of successfully receiving $\ell$ descriptions, then in order to guarantee stability, the average quantization noise variance needs to satisfy:
\begin{equation}\label{eq:avg_var}
\sum_{\ell = 0}^{k} p_s(\ell) \sigma^2(\ell)  < \sigma_v^2 \gamma^{-1},
\end{equation}
 where $\sigma^2(\ell)$  is the equivalent noise variance when $\ell$ descriptions are received.
 \end{corollary}

\section{Simulation Study}\label{sec:sim}

\subsection{Single Description Setup}
\begin{figure}[th]
\begin{center}
\includegraphics[width=8cm]{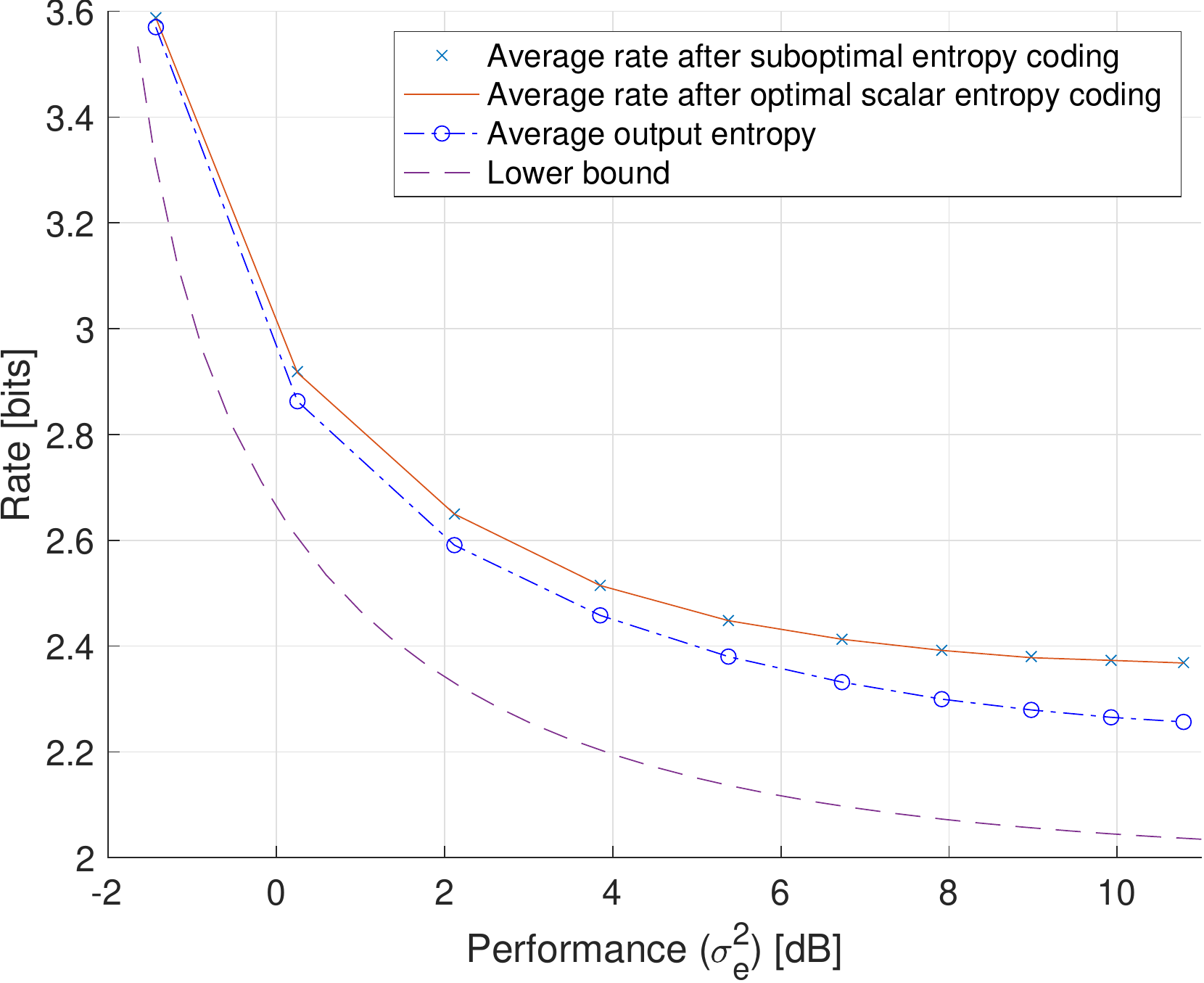}
\caption{ The control performance of stabilizing and repetition codes as a function of packet-loss probability.}
\label{fig:performance_sd}
\end{center}
\vspace{-5mm}
\end{figure}

We first consider a single-description setup, where a plant with the following input-output relationship between $(u,d)$ and $y$ is  controlled over an errorless digital channel:
\begin{equation}\label{eq:io2}
y = \frac{0.165}{(z-4)(z-0.5789)}( u + d), \quad e=y.
\end{equation}
Since the system has a single unstable pole at $z=4$, the minimum rate for stability is $\mathcal{R} = 0.5\log_2(4) = 2$ bits/sample. In Fig.~\ref{fig:performance_sd}, we have plotted the lower bound of~\cite{silva:2016} describing the minimum data rate in bits/sample required to stabilize the system and to achieve a desired performance (output variance $\sigma_e^2$). The performance is shown in dB ($10\log_{10}(\sigma_e^2)$). We have obtained the filters $F,L$ using the method outlined in \cite{silva:2016}, and simulated the dynamical system using $M=10^6$ samples. We have used a scalar quantizer $\mathcal{Q}_\Delta$, where we have varied the step-size $\Delta = \sqrt{12}n, n=1,\dotsc, 10$. The resulting data rate and performance for each choice of $\Delta$ is  also shown in Fig.~\ref{fig:performance_sd}. 

The output variable $v$ of the filter $L$ is quantized using $\mathcal{Q}_\Delta$. 
The  average  entropy of $\mathcal{Q}_\Delta(v)$ is represented by the dash-dotted curve labeled "average output entropy". This is the empirical entropy obtained by replacing the probabilities by relative frequency counts. Specifically, let $\#a$  be the number of occurences of the  symbol $a\in \Delta \mathbb{Z}$. Then the relative frequency count is $p_a = \#a/M$, and the average empirical entropy is $-\sum_{a\in \Delta\mathbb{Z}} p_a \log_2(p_a)$. 

We have also shown the average data rate after entropy coding using an optimal scalar Huffman entropy coder, which is designed with knowledge of the realizations of $\mathcal{Q}_\Delta(v)$. Thus, this is the ideal case where one would know the exact distribution when designing the entropy coder. 
For comparison, we have shown the operational average data rate obtained by using a sub optimal entropy coder (blue crosses in Fig.~\ref{fig:performance_sd}), which is designed using a quantized (with the correct step-size) Gaussian distribution with a variance that is identical to that of $w$. This means that the entropy coder is designed with only limited knowledge about the distribution, i.e., only knowledge of the variance. 
It can be noticed that the operational data rates using a "Gaussian" entropy coder coincides with the data rate obtained using knowledge of the ideal statistics. Thus, even though the encoder is unaware of the particular erasures, there is practically no loss by assuming a Gaussian distribution. 
In the remaining simulations we therefore simply use these operational average data rates after optimal scalar entropy coding as a good indication of the practical data rates of the system. 

In Fig.~\ref{fig:performance_sd}, the lower bound represents the theoretically achievable performance obtainable if one could generate Gaussian quantization noise and if there was no loss of the entropy coder. It is interesting to note that the average entropy is approximately 0.25 bits greater than the lower bound, which corresponds to divergence between a Gaussian and a uniform distribution.  It can also be observed that the operational data rates obtained when using an entropy coder are less than 0.15 bits greater than the entropy. Thus, the operational data rates obtained using a scalar quantizer followed by sub optimal entropy coding is only about 0.4 bits/sample away from the lower bound. 

\subsection{Independent Encodings}

In this simulation we design a $(2,1)$ and a $(3,2)$ stabilizing code based on independent encodings. We consider the same system in \eqref{eq:io2} as in the previous simulation.
We use a subtractively dithered scalar quantizer, which is a stochastic quantizer that provides different outputs, when encoding the same source multiple times \cite{zamir:2014}. We use this to form the $k$ \emph{independent} encodings.
\begin{figure}[th]
\begin{center}
\includegraphics[width=8.5cm]{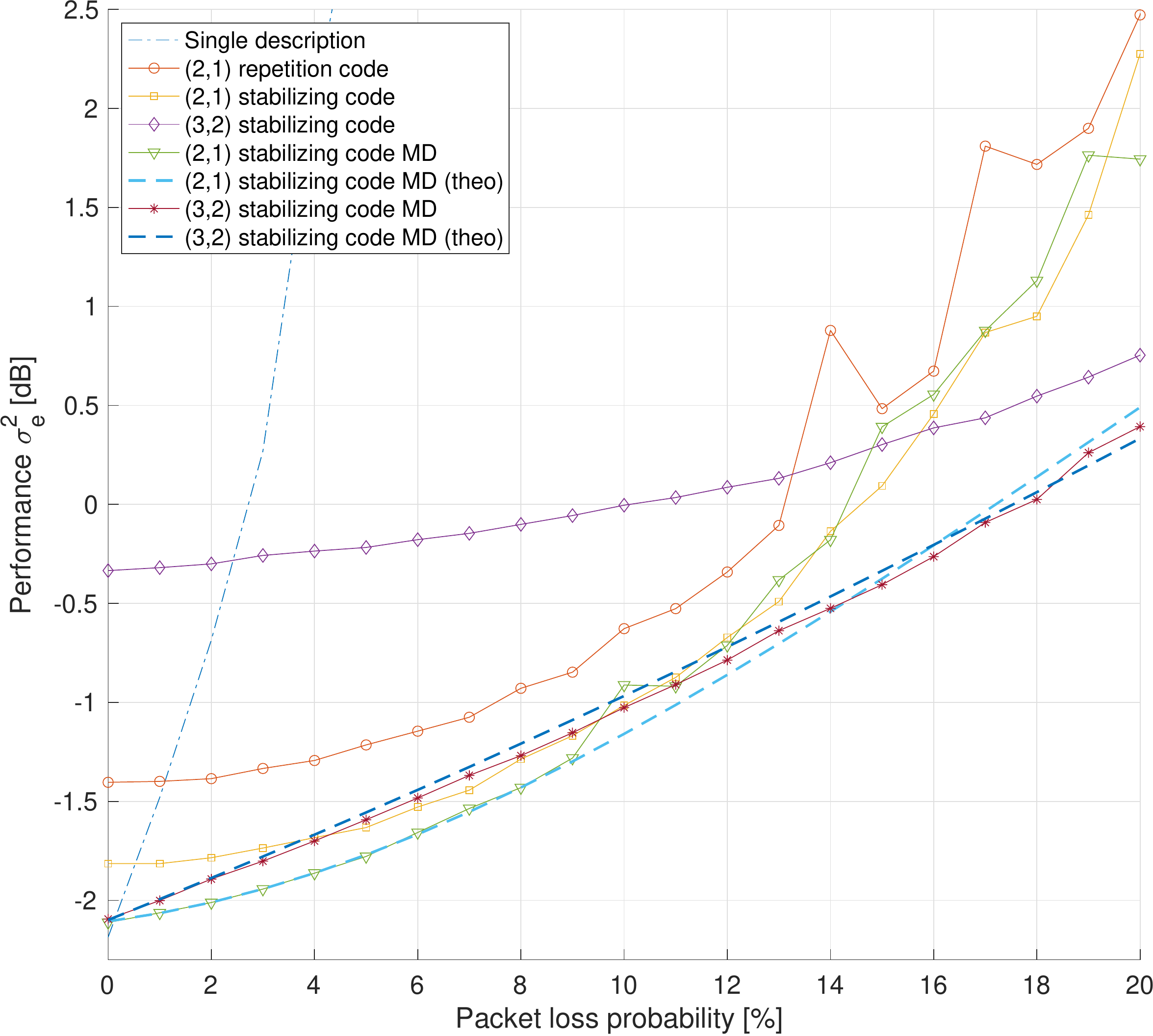}
\caption{ The performance of stabilizing and repetition codes as a function of packet-loss probabilities in the range $[0, 20\%]$.}
\label{fig:performance_all_zoom}
\end{center}
\vspace{-5mm}
\end{figure}

\begin{figure}[th]
\begin{center}
\includegraphics[width=8.5cm]{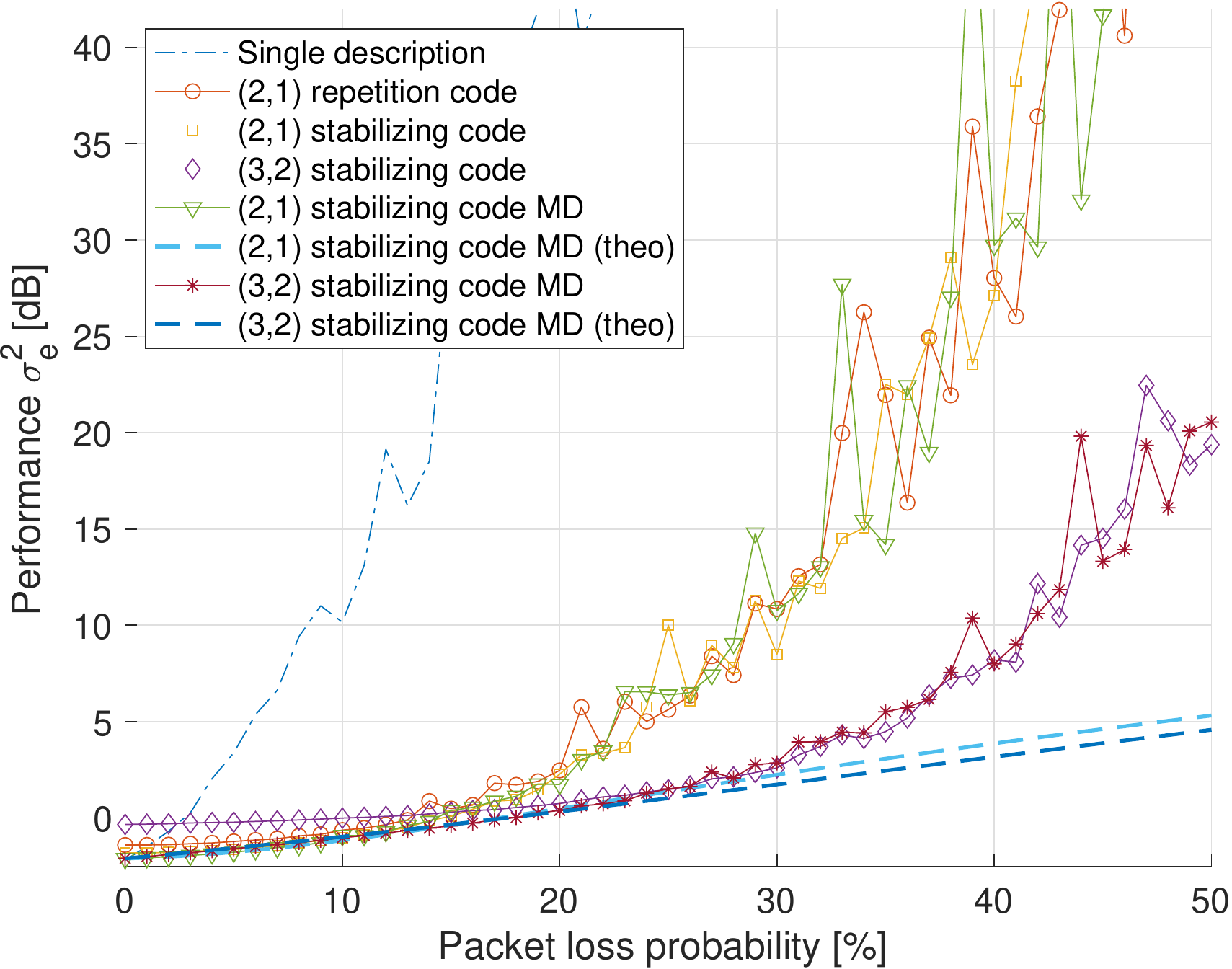}
\caption{ The performance of stabilizing and repetition codes as a function of packet-loss probability.}
\label{fig:performance_all}
\end{center}
\vspace{-5mm}
\end{figure}


We encode the output $v$ of Fig.~\ref{fig:system} (right) using a  subtractively dithered scalar quantizer $\mathcal{Q}_\Delta$ with step-size $\Delta = 12$, which implies that the resulting quantization noise variance is $\Delta^2/12 = 12$. We design the system $(L,F)$ such that $\sigma_v^2 = 133$, which implies that the resulting $\mathrm{SNR}$ when using a single description is $133/12 \approx 11$, and when combining two descriptions we obtain an $\mathrm{SNR}$ of $2\cdot 133/12 \approx 22$. Thus, the system is not stable when receiving a single description but stable when receiving at least two descriptions. 

We have plotted the performance of the $(3,2)$ stabilizing code in Figs.~\ref{fig:performance_all_zoom} and~\ref{fig:performance_all}  as a function of the packet-loss probability. We assume i.i.d.\ packet losses, and simply average the received descriptions to form the reconstruction --- except when all three descriptions are received in which case we can use the inverse map $\phi^{-1}$. 

In order to design a $(2,1)$ stabilizing code, we  choose $\Delta = 4$. The resulting $\mathrm{SNR}$ when receiving a single description is then $133/(\Delta^2/12) \approx 100$, which is much greater than 15 and the system is therefore stable even when receiving a single description. The choice of $\Delta$ is made such that the average sum-rates are equivalent for the $(2,1)$ and $(3,1)$ stabilizing codes and given by approximately $7.1$ bits/sample. The performance of this code is also shown in Figs.~\ref{fig:performance_all_zoom} and~\ref{fig:performance_all}. 
The performance when transmitting one of the descriptions two times is also shown in Figs.~\ref{fig:performance_all_zoom} and~\ref{fig:performance_all}. This corresponds to a $(2,1)$ repetition code having similar sum-rate as the $(2,1)$ and $(3,2)$ stabilizing codes. For comparison, we have also plotted the performance of a single--description system with a rate of 7 bits/sample.
For each packet-loss probability, the performance and rates are averages over a realization having $10^6$ samples. In the case where two descriptions are received, the stabilizing code is able to improve its performance beyond what is possible when using only a single description. This is in constrast to the repetition code, where both descriptions share the exact same information and no gain is possible when receiving both descriptions. It can therefore be observed that a low packet loss probabilities, the performance of the stabilizing code is better than that of the repetition code.

The stabilizing code based on multiple descriptions further improve upon the code based on independent encodings as is shown in Section~\ref{sec:sim_md}. This is possible since the quantization noises becomes negatively correlated when using multiple descriptions, whereas they are not correlated for the independent encodings.

\subsection{Multiple Descriptions}

\subsubsection{Performance using fixed subsets of descriptions}
In this simulation we demonstrate the rate and control performance when replacing the scalar quantizer by a multiple description scalar quantizer. 
We use the multiple description index-assignment method of \cite{ostergaard:2006} and we use $k=3$ descriptions. We consider the case of no packet losses, and design a $(3,1)$ stabilizing code, which means that the system is stable even if we only use a single description out of the $k=3$ descriptions. 
We choose a nesting ratio of $\Delta_s/\Delta = 7$. The normalized index-assignment map $\phi$ for the 49 central quantizer points that are closest to the origin are shown in Table~\ref{tab:phi1}.
The first column contains the central quantizer points, and the other 3 columns contains the $k$-tuples assigned to the central quantizer points. For example, the third row in Table~\ref{tab:phi1} (left), shows that the central quantizer point $b=2$ is mapped to $a_1=7, a_2=0,$ and $a_3=0$. All the entries in the table are normalized so that they assume that the step-size of the central quantizer is $\Delta = 1$. To use an arbitrary step-size $\Delta \in \mathbb{R}_+$ one  has to multiply all entries in the tables by $\Delta$. In order to obtain the complete mappings for all elements in the central quantizer, one has to tesselate the space by translating the entries in the table by adding multiples of $49$. For example, to find the mapping for $b=27$ we notice that $27 - 49 = -22$. Thus, we use the mapping for $b=-22$ which is $(-14,-21,-28)$ and simply add 49 to obtain the mapping for $b=27$, which is $(35,28,21)$. 

\begin{table}[th]
\begin{center}
\caption{Normalized index-assignment map $\phi$ for $k=3$ and $\frac{\Delta_s}{\Delta}=7$. }
\begin{tabular}{|p{1mm}|p{1mm}|p{1mm}|p{1mm}|}\hline
$b$ & $a_1$ &  $a_2$ &  $a_3$ \\ \hline
     0 &    0 &    0&     0 \\
     1  &   0  &   0 &    7\\
     2  &   7  &   0&     0\\
     3 &    0 &    7&     0 \\
     4 &    0 &    7&     7\\
     5 &    7  &   7&     0\\
     6 &    7  &   0&     7\\
     7  &   7  &   7 &    7\\
     8 &   14 &    7&     7\\
     9 &    7  &  14&     7\\
    10 &    7 &    7&    14\\
    11 &    7  &  14&    14\\
    12 &   14&    14&     7\\
    13 &   14&     7 &   14\\
    14 &   14&    14&    14\\
    15 &   14 &   14&    21\\
    16 &   21  &  14&    14\\ \hline
\end{tabular} 
\begin{tabular}{|p{4mm}|p{4mm}|p{4mm}|p{4mm}|}\hline
$b$ & $a_1$ &  $a_2$ &  $a_3$ \\ \hline
    17   & 14  &  21&    14\\ 
    18 &   14 &   21&    21\\
    19 &   21  &  21&    14\\
    20 &   21 &   14&    21\\
    21 &   21  &  14&    28\\
    22 &   14 &   21 &   28\\
    23 &   21 &   21 &   21\\
    24&    21 &   21 &   28\\ 
-24    & -21  & -21  & -28 \\
   -23 &  -21 &  -14 &  -28\\
   -22 &  -14 &  -21 &  -28\\
   -21  & -21 &  -21  & -21\\
   -20 &  -21 &  -21 &  -14\\
   -19  & -21 &  -14 &  -21\\
   -18  & -14 &  -21 &  -21\\
   -17  & -14 &  -14 &  -21\\
   -16  & -21 &  -14 &  -14\\ \hline
\end{tabular}
\begin{tabular}{|p{4mm}|p{4mm}|p{4mm}|p{4mm}|}\hline
$b$ & $a_1$ &  $a_2$ &  $a_3$ \\ \hline
   -15  & -14 &  -21 &  -14\\ 
   -14 &  -14 &  -14 &  -14\\ 
  -13  & -14 &  -14 &   -7\\
   -12  & -14  &  -7  & -14\\
   -11  &  -7 &  -14   &-14\\
   -10  &  -7&    -7   &-14\\
    -9  & -14&    -7   & -7\\
    -8  &  -7 &  -14   & -7\\
    -7  &  -7 &   -7   & -7\\
    -6  &  -7 &   -7   &  0\\
    -5  &  -7&     0   & -7\\
    -4  &   0&    -7   & -7\\
    -3  &   0&     0    &-7\\
    -2  &  -7&     0   &  0\\
    -1   &  0&    -7    & 0\\ \hline
\end{tabular}
\label{tab:phi1}
\end{center}
\vspace{-5mm}
\end{table}

\begin{figure}[th]
\begin{center}
\includegraphics[width=8cm]{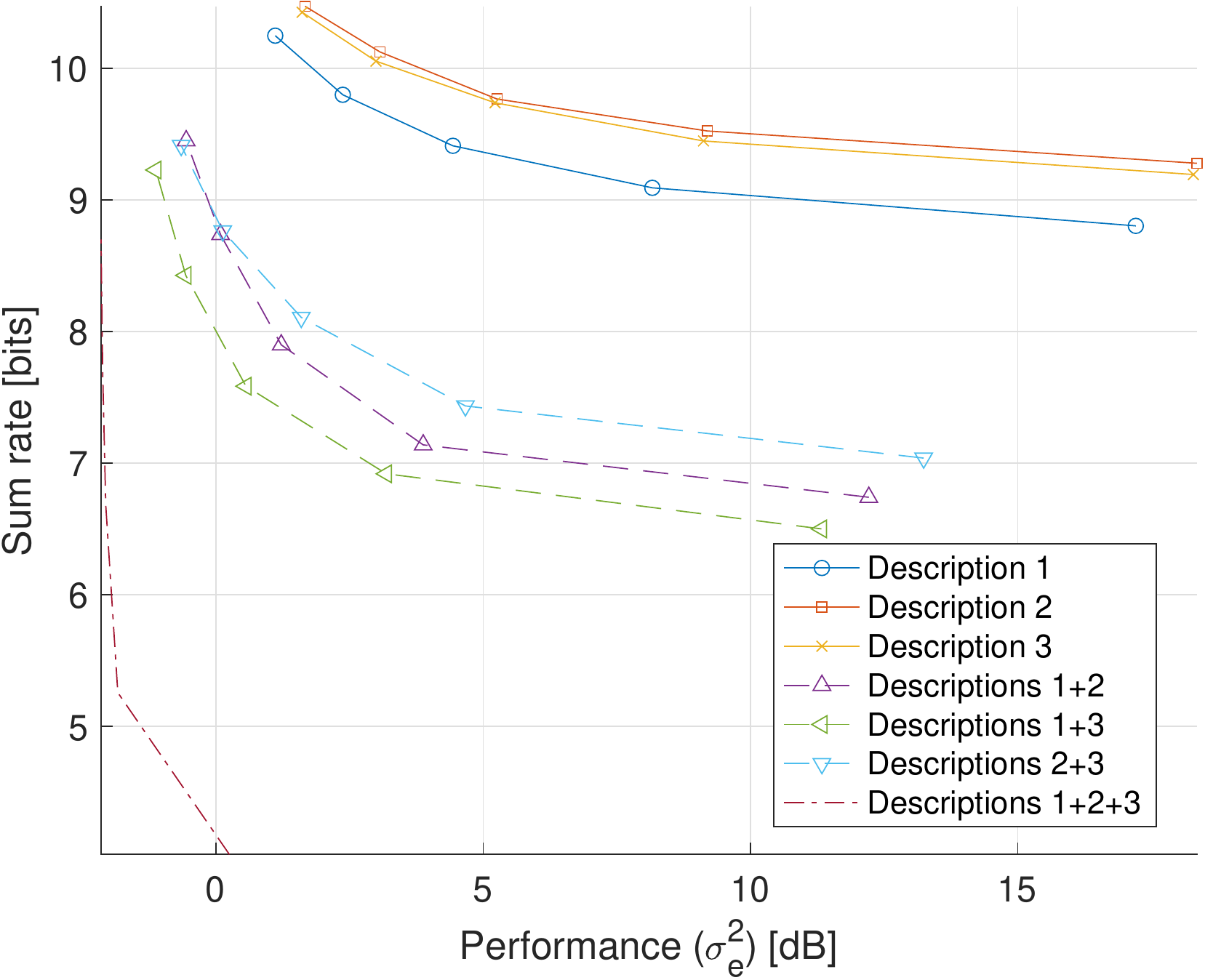}
\caption{Rate and control performance curves for multiple descriptions with a nesting ratio of $\Delta_s/\Delta = 7$. The step size $\Delta$ is varied in the interval $2\sqrt{12}/n, n=1,3,5,7,9$.}
\label{fig:performance_md_N7}
\end{center}
\vspace{-5mm}
\end{figure}

In Fig.~\ref{fig:performance_md_N7} we have shown the performance when we always only use the first description out of the $k=3$ descriptions (solid line with circles). This means that even though we form three descriptions, the control signal applied to the plant is only based upon the first of the three descriptions. The system is stable and we can measure the average data rates after optimal scalar entropy coding. Since we are transmitting three descriptions we report the sum-rate over all three descriptons (even though we are only using one out of three descriptions). Similarly, we have in Fig.~\ref{fig:performance_md_N7} shown the performance when using only description 2, and when only using description 3. 

As described in Section~\ref{sec:md} it is possible to combine descriptions to get improved performance. Let us assume that we  always only receive descriptions 1 and 2. We then combine them by forming their average. The system is of course also stable in this case, and the performance is significantly improved as can be seen in Fig.~\ref{fig:performance_md_N7}. We have similarly shown the performance when combining descriptions 1 and 3 as well as when combining descriptions 2 and 3. We observe that for a fixed performance, the sum-rate can be significantly reduced by using 2 out of 3 descriptions compared to what it is possible when only using a single description. For comparison we also show the performance when we always use three descriptions, which provides further improvement over that what is possible when only using two descriptions. Recall that when using all three descriptions, we basically recover the situation where we are using a single description scalar quantizer. However, comparing the performance of using all three descriptions to that obtained using a single-description system as shown in Fig.~\ref{fig:performance_sd}, we notice that it is not as good. The reason is that splitting the information into three partially redundant descriptions leads to an increased data rate compared to using a single description. Of course if we have packet losses, it is better to use more descriptions. 

The numerically obtained average estimates $D_\ell$ of the variances of the quantization errors are shown in Table~\ref{tab:distortions} for the case of $\ell=1,2,3$. 
For comparison, we have  also shown the theoretical quantization noise variance $\sigma^2(\ell)$ approximation given by \eqref{eq:sigmaell}. It can observed that the approximations are quite close to the numerically obtained estimates. 
\begin{table}[th]
\begin{center}
\caption{Practical $D_\ell$ and theoretical  $\sigma^2(\ell)$  quantization noise variance (in dB) due to combining $\ell=1,2,3$ descriptions. }
\begin{tabular}{c|llllll} 
$\Delta$  &  $D_1$ &  $\sigma^2(1)$  & $D_2$ &   $\sigma^2(2)$  & $D_3$  & $\sigma^2(3)$ \\ \hline
$2\sqrt{12}$ &27.38   & 27.84  & 21.65  & 21.91  &   6.02   & 6.02  \\
$2\sqrt{12}/3$ &   17.94  & 18.30 &  12.20 &  12.37 &  -3.51  & -3.52  \\
$2\sqrt{12}/5$  &  13.51  & 13.87  &  7.95  &  7.93  & -7.96   &-7.96  \\
 $2\sqrt{12}/7$ &  10.64  & 10.94  &  5.25 &   5.01 & -10.89 & -10.88  \\
 $2\sqrt{12}/9$  &  8.51  &  8.76   & 3.22  &  2.82 & -13.07 & -13.06 \\ \hline
\end{tabular}
\label{tab:distortions}
\end{center}
\vspace{-5mm}
\end{table}

%


\subsubsection{Control performance using multiple descriptions}\label{sec:sim_md}
%
We now demonstrate the control performance of the system when using stabilizing codes based on multiple descriptions and when the channel has i.i.d.\ packet losses. 
Towards that end, we show how to design $(2,1)$ and $(3,2)$ stabilizing codes. 

In principle 
given a desired average control performance and the packet loss rate of the channel, one could perform a joint design of the encoder+decoder (which includes the controller)  and then the erasure code by selecting 
 $\gamma$ (and thereby also $F,L,P$) and $\Delta_s,\Delta, k, k'$ so as to minimize the total sum rate. 
 
 In this example, we choose $\Delta_s/\Delta=7$ which yields the map $\phi$ shown in Table~\ref{tab:phi1}.  
Due to the unstable pole at $z=4$, the minimal coding rate for a single-description system to guarantee stability is $\mathcal{R}_s \geq \log_2(4)=2$ bits/sample. Since the lower bound on the coding rate can also be expressed as $\frac{1}{2}\log_2(1 + \mathrm{SNR})\geq 2$ it follows that we require $\mathrm{SNR} > 2^{4}-1 = 15$. 

We first construct the $(3,2)$ stabilizing code. Thus, we wish to design a code that is stable when receiving at least $\ell = 2$ descriptions, and we therefore need to guarantee that $\mathrm{SNR} = \sigma_v^2/\sigma^2(\ell) > 15$, where $\sigma^2(\ell)$ is given by \eqref{eq:sigmaell} and also in Table \ref{tab:distortions} for selected $\Delta$ values. Let us choose the step-size of the central quantizer as $\Delta=2\sqrt{12}/5$. Then from Table~\ref{tab:distortions}, we observe that $\sigma^2(2) \approx 6.3$ (8 dB). Thus, we require $\sigma_v^2>6.3 \cdot 15 =94.5$. In order to achieve a certain performance one needs to use a greater $\mathrm{SNR}$ than the lower bound. In this case, we choose to design the filters $(L,F)$ using the variance $\sigma_v^2 = 133$. This yields  $\mathrm{SNR} = 133/6.3 \approx 22$, which is strictly greater than the lower bound of $15$. On the other hand $\sigma_v^2/\sigma^2(1) = 133/25 \approx 5$, which is below the lower bound. The  code is therefore not stabilizing when only receiving a single description. The performance is shown in Fig.~\ref{fig:performance_all}.  

If we use \eqref{eq:avg_var}, then theoretically for packet loss probabilities above 37\%, the system becomes unstable. Moreover, based on the MJLS results, the system becomes unstable if the largest absolute eigenvalue of $\mathbb{A}$ in \eqref{eq:AA} is greater than 1.  We find that the largest absolute eigenvalue of $\mathbb{A}$ is $0.8785, 0.9496, 1.0245$ for packet loss probabilities of $38\%, 39\%$, and $40\%$, respectively. Indeed, we observe in practice that the system is not always stable for packet loss probabilities above 40\%. 

We now construct the $(2,1)$ stabilizing code. Then, since $k=2$ and $\ell=1$, it follows from \eqref{eq:sigmaell} that:
\begin{equation}
\sigma^2(1) = \frac{\Delta^2}{12} + \frac{1}{4}\frac{\Delta^2}{12} \bigg(\frac{\Delta_s}{\Delta}\bigg)^{4}.
\end{equation}
We require that $\sigma_v^2/\sigma^2(\ell) > 15$ for stability. Using the same $\Delta=1.33$ and $\Delta_s/\Delta =3$, we obtain $\sigma^2(1) = 3.14$. We choose $\gamma=133$ as before, which yields $133/3.14 \approx 42$, which is much greater than $15$. The code is therefore a $(2,1)$ stabilizing code. The index-assignment map $\phi$ for $k=2$ and $\Delta_s/\Delta=3$ is shown in Table~\ref{tab:phi4} (left). For comparison, we have also shown the index-assignment map $\phi$ for $k=3$ and $\Delta_s/\Delta=3$ in Table~\ref{tab:phi4} (right).

\begin{table}[tb]
\begin{center}
\caption{Normalized index-assignment map $\phi$ for $k=2$ (left) and $k=3$ (right)  for $\Delta_s/\Delta=3$.}
\begin{tabular}{|p{2.5mm}|p{2.5mm}|p{2.5mm}|p{2.5mm}|} \hline
$b$ & $a_1$ &  $a_2$  \\ \hline
 -4    &-6   & -3 \\
    -3 &   -3  &  -3\\
    -2 &   -3  &   0\\
    -1  &   0   & -3\\
     0  &   0   &  0\\
     1   &  3    & 0\\
     2  &   0   &  3\\
     3  &   3   &  3\\
     4  &   6   &  3\\ \hline
\end{tabular} 
\begin{tabular}{|p{2.5mm}|p{2.5mm}|p{2.5mm}|p{2.5mm}|p{2.5mm}|} \hline
$b$ & $a_1$ &  $a_2$ &  $a_3$ \\ \hline
-4   & -3   & -3  &  -6   \\
    -3 &   -3   & -3 &   -3\\
    -2  &   0   & -3  &  -3\\
    -1    & 0   &  0   & -3\\
     0    & 0   &  0   &  0\\
     1    & 0   &  0    & 3\\
     2   &  0   &  3   &  3\\
     3    & 3  &   3   &  3\\
     4    & 3 &    3    & 6\\ \hline
\end{tabular}
\label{tab:phi4}
\end{center}
\vspace{-5mm}
\end{table}

%

It can be seen in Fig.~\ref{fig:performance_all} that stabilizing codes outperforms repetition coding. Moreover, using multiple descriptions when constructing the stabilizing codes is better than using independent encodings, except at very low bitrates or very high packet-loss rates, where they achieve nearly identical performance. The theoretical performance of the stabilizing codes based on multiple descriptions can be assessed in the following way. First, from the left hand side of \eqref{eq:avg_var}, the average variance of the quantization noise $\sigma_q^2$ can be obtained. The control performance is then obtained by inserting into 
\eqref{eq:perf}. We have plotted the theoretical achievable performance for the case of $k=2,3$ in Figs.~\ref{fig:performance_all_zoom} and~\ref{fig:performance_all}. For low to moderate packet loss probabilities we observe in Fig.~\ref{fig:performance_all_zoom} a very good correspondence between the theoretically and numerically obtained performances. On the other hand, from Fig.~\ref{fig:performance_all} it can be seen that for higher packet loss probabilities, the theoretical predictions of the performances are no longer valid. In this region it becomes harder to control the system due to an increasing amount of packet dropouts.
We note that all coding schemes are designed so that they use the same total transmit rate. The measured operational rates are provided in Table~\ref{tab:eff}. 

\subsection{Practical Efficiency}
The practical efficiency can be assessed as:
\begin{equation}\label{eq:prac_eff}
\eta = \frac{ \frac{1}{2}\log_2(1 + 12\sigma_v^2/\Delta^2) + \frac{1}{2}\log_2(2\pi/6)}
{\mathcal{R}_s},
\end{equation}
where $\mathcal{R}_s$ denotes the measured average sum-rate after optimal \emph{scalar} entropy coding. In theory, repetition coding has a practical efficiency of $1/k$ since we are transmitting the same information $k$ times. However, in practice there is a gap due to the loss of the \emph{scalar} quantizer and entropy coder, since these losses are included $k$ times.  

We have measured $\sigma_v^2 \approx 120$ in this simulation, which when using \eqref{eq:Rs_mdc} leads to a theoretical sum-rate $\mathcal{R}_s = 0.5 k\log_2(2\pi e \sigma_v^2) - k\log_2(\Delta_s)   = 6.67$ bits/sample, which yields a data rate per description of approximately $2.3$ bits/sample. 
The operational sum-rate measured in practice is 7.12 bits/sample, which is close to the theoretical value. The rate required to achieve the performance of the central quantizer is $\mathcal{R} = 0.5\log_2( 1+ 12\sigma_v^2/\Delta^2) - 0.5\log_2(\pi e /6) \approx 4.51$, which yields a practical efficiency of $4.51/7.12 \approx 0.63$ when using \eqref{eq:prac_eff}. 

The measured data rates and practical efficiency \eqref{eq:prac_eff} of the stabilizing codes are shown in Table~\ref{tab:eff}. Also shown are the theoretical sum-rates \eqref{eq:Rs_mdc}. It can be seen that the numerically obtained rates are close to the theoretical values. It can also be observed that the practical efficiency of stabilizing codes based on multiple descriptions (MD) is significantly better than designs based on independent encodings or repetition codes. 
\begin{table}[th]
\begin{center}
\caption{Measured average data rate $\mathcal{R}_s$, theoretical rate \eqref{eq:Rs_mdc}, and practical efficiency $\eta$ of the proposed stabilizing codes.}
\begin{tabular}{lllll}
Code type &  $\eta$ & $\mathcal{R}_s$ & \eqref{eq:Rs_mdc} \\ \hline
$(2,1)$ repetition code      & 0.42 & 7.14 & ---\\ 
$(2,1)$ stabilizing code &  0.43 & 7.16 & --- \\
$(2,1)$ stabilizing code MD &  0.65 & 7.08 & 7.00  \\
$(3,1)$ repetion code&   0.25   & 7.11 &  ---\\
$(3,2)$ stabilizing code  & 0.25  & 7.07 &  --- \\
$(3,2)$ stabilizing code MD & 0.63 & 7.12 & 6.65  \\ \hline
\end{tabular}
\label{tab:eff}
\end{center}
\vspace{-5mm}
\end{table}

\section{Conclusions}
A new construction of delayless error correction codes for stabilizing systems over erasure channels were proposed. The proposed codes take the packet loss probabilities of the channel and the stability of the system into account. 
They are simple to use in practice. For the encoder, it is a scalar quantizer followed by a look-up table. For the decoder, you either average the received decoded descriptions or use the inverse map (look-up table) in case you receive all descriptions. 

Conventional erasure correction codes add redundant information such as parity bits in order to be robust to erasures. The parity bits do not contain any useful information about the control signals and are wasteful in the case of none or only few erasures.  On the other hand, a distinguishing aspect of the proposed codes is that the added redundancy contains useful  information about the control signals. Specifically,  
each source packet is encoded into $k$ channel packets in such a way that receiving $k'$ channel packets, the decoder is instantaneously able to reconstruct a control signal of sufficient quality to guarantee stability. Moreover, if more than $k'$ packets are received, then a more accurate control signal leading to a better performance of the system is obtained. This is not possible with classical erasure correction codes. Moreover, if less then $k'$ channel packets are received, some control signal information can still be recovered, however, there is not sufficient information for guaranteeing stability. 

One can also design the stabilizing erasure code to guarantee stability as long as that on average at least $k'$ packets are received. 
For linear systems with scalar input and output, explicit designs were provided, and a simulation study with i.i.d.\ erasures revealed that there is a significant gain over using traditional repetition codes. 
Similar to repetition coding, the proposed codes do not add additional delays.


%

\bibliography{literature} 
\bibliographystyle{IEEEtran}
%
%
%


\appendix

\section{Proofs of Lemmas}
\begin{proof}[\textbf{Lemma~\ref{lem:indp}]}
The variance of $\frac{1}{k'}(y_{i_1} + \cdots + y_{i_{k'}})$ is $(k')^{-1}\sigma^2$ for any subset of $k'$ encodings. The resulting $\mathrm{SNR}$, say $\gamma' = k' \sigma_v^2 \sigma^{-2}$, when combining $k'$ descriptions needs to satisfy:
\begin{align} \label{eq:1}
\gamma' = \frac{k'\sigma_v^2}{\sigma^2} > \|S-1\|^2,
\end{align}
since $\|S-1\|^2$ is the minimal $\mathrm{SNR}$ required to guarantee stability. We now use that $\gamma \sigma_q^2= \sigma_v^2$, and from \eqref{eq:gamma} we get:
\begin{align}\label{eq:sv}
 \sigma_v^2 =  \gamma (\gamma - \|S-1\|^2)^{-1}\|L_yP_{21}S\|^{2}.
\end{align}
Inserting into \eqref{eq:1} and re-arranging terms leads to:
\begin{align}\label{eq:sn2}
\sigma^2 < \gamma k' \|S-1\|^{-2} (\gamma - \|S-1\|^2)^{-1}\|L_yP_{21}S\|^{2},
\end{align}
which leads to \eqref{eq:code1}. 
\end{proof}

\begin{proof}[\textbf{Lemma~\ref{lem:R}]}
Let $\sigma^2$ be the variance of the coding noise for a single description of the $(k,k')$ stabilizing code. Then, the resulting variance when linearly combining $k'$ descriptions is $\sigma^2/k'$. Thus, 
$\mathrm{SNR} = k\sigma_v^2/ \sigma^2 \geq  \|S-1\|^2$, where the inequality follows since $\|S-1\|^2$ is the minimum $\mathrm{SNR}$ that guarantees stability. Isolating $\sigma^2$ leads to:
\begin{equation}\label{eq:s2lb}
\sigma^2 \leq \|S-1\|^{-2} k' \sigma_v^2. 
\end{equation}
We can now express the \emph{ideal} sum-rate in terms of $\sigma^2$:
\begin{align*}
\mathcal{R}_s = \frac{k}{2}\log_2(1 + \frac{\sigma_v^2}{\sigma^2}) 
\geq \frac{k}{2}\log_2(1 + (k')^{-1}\,\|S-1\|^{2}).
\end{align*}
\end{proof}

\begin{proof}[\textbf{Lemma~\ref{lem:eta_indp}]}
The first part follows immediately from \eqref{eq:s2lb}, since the $\mathrm{SNR}$ for a single description is $\sigma_v^2/\sigma^2$ and for $k$ descriptions it is $ k\sigma_v^2/\sigma^2$. The second part 
  follows since the logarithm of the number $1+kc$ is approximately linear in $k$ when $c\ll 1$, i.e., $\log(1+kc) \approx k\log(1+c)$ for small $c$. 
\end{proof}

\begin{proof}[\textbf{Lemma~\ref{lem:indp4}]}
Follows from \eqref{eq:perf} by inserting \eqref{eq:sn2} and the fact that the noise variance satisfies $\frac{\sigma^2}{\ell}$ for $\ell=1,\dotsc, k$. 
\end{proof}

\begin{proof}[\textbf{Lemma~\ref{lem:md}]}
We need to ensure that $\mathrm{SNR}>\|S-1\|^2$, when receiving at least $k'$ descriptions. 
The  noise variance when combining any $k'$ descriptions is given by:
\begin{align}\label{eq:var}
\mathrm{var}\bigg( \frac{1}{k'} \sum_{i=1}^{k'} q_i \bigg) = \frac{\sigma^2}{k'}(1+(k'-1)\rho).
\end{align}
Using \eqref{eq:var},  the $\mathrm{SNR}$ is given by:
\begin{align}
\frac{\sigma_v^2}{\frac{\sigma^2}{k'}(1+(k'-1)\rho)} \geq \|S-1\|^2.
\end{align}
Isolating $\sigma^2$ and inserting \eqref{eq:sv} leads to:
\begin{align}
\sigma^2 &\leq k'\|S-1\|^{-2}(1+(k'-1)\rho)^{-1}\sigma_v^{-2} \\ \notag
&= k'\|S-1\|^{-2}(1+(k'-1)\rho)^{-1}
\gamma (\gamma - \|S-1\|^2)^{-1} \\ \notag
&\quad\times \|L_yP_{21}S\|^{2}.
\end{align}
\end{proof}

\begin{proof}[\textbf{Lemma~\ref{lem:prac}]}
At high-resolution conditions, the rate of a single description system using scalar quantizer is $\mathcal{R} = 0.5\log_2(1 + \sigma_v^2/\sigma_q^2) + 0.5\log_2(2\pi/6)$, since the space-filling loss of a scalar quantizer becomes exactly $0.5\log_2(2\pi/6)$  and the loss of the entropy coder dissappears \cite{linder:2006}. Note that $0.5\log_2(2\pi/6)\approx 0.25$ bits/sample so the practical efficiency differs only slighty from the effiency defined without including the space-filling loss. 
Recall that when receving all $k$ descriptions, the performance of the single descriptions quantizer with a step-size of $\Delta$ is achieved using the rate given in 
 \eqref{eq:Rs_mdc}. Let us compare the ratio $\mathcal{R}/\mathcal{R}_s$:
\begin{align}\notag
 \lim_{\Delta \to 0} \frac{\mathcal{R}}{\mathcal{R}_s} &=  \lim_{\Delta \to 0} \frac{\frac{1}{2}\log_2(1 + \sigma_v^2/\sigma_q^2) + \frac{1}{2}\log_2(2\pi/6)}{\frac{k}{2}\log_2(2\pi e \sigma_v^2) - k\log_2(\Delta_s)} \\ \notag
&=  \lim_{\Delta \to 0} \frac{\frac{1}{2}\log_2(1 + 12\sigma_v^2/\Delta^2) + \frac{1}{2}\log_2(2\pi/6)}{\frac{k}{2}\log_2(2\pi e \sigma_v^2) - k\log_2(\Delta_s)} \\ \label{eq:eff_mdc}
&= \lim_{\Delta \to 0}  \frac{\log_2(\Delta)}{k\log_2(\Delta_s)} = \frac{1}{k},
\end{align}
which shows that the efficiency of the proposed construction is asymptotically (in rate)  no better than that of a repetition code.
\end{proof}

\begin{proof}[\textbf{Lemma~\ref{lem:mjls}]}
We clearly have a finite set of possible jump states (and decoders), which is upper bounded by $2^k$. 
Since we are constructing  symmetric descriptions, the reconstruction quality only depends upon the number of received descriptions and not which descriptions that are received. This means that while there are $2^k$ possible decoders, there are only $(k+1)$ possible reconstruction qualities. We express the reconstruction quality implicitly via the variance $\sigma_q^2$ of the quantization noise. Let $\xi_i \in \{0,\dotsc, k\}$ be the \emph{jump} variable that indicates the index of the quality of the reconstruction at time $i$. Specifically, $\xi_i = |\mathcal{I}(i)|$, i.e., the cardinality of the set of indices of the received descriptions at time $i$.

Let $M(\xi_i) \in \{0,1\}$ be the indicator function defined as:
\begin{equation}
M(\xi_i) = \begin{cases}
0, & \xi_i =0, \\
1, & \text{otherwise}. 
\end{cases}
\end{equation}
and let the vector $\bar{q}(i) \in \mathbb{R}^{k+1}$ of quantization noises be such that
\begin{equation}\label{eq:barq}
\bar{q}(i) = [0, q_1, q_2,\dotsc, q_k],
\end{equation}
and $\mathrm{var}(q_{\xi_i}) = \sigma^2(\xi_i)$ for $\xi\geq 1$, where $ \sigma^2(\xi_i)$ is given by \eqref{eq:sigmaell}. Finally, let $J(\xi)\in\mathbb{R}^{k+1}$ be the all-zeros vector except that it has a one as its $(\xi+1)th$ entry. For example, if $k=4$ and $\xi_i=2$, then $J=[0,0,1,0,0]^T$.
With this notation, we can write the output $w(i)$ of the quantizer at time $i$ as a function of the \emph{jump} variable $\xi_i$, that is:
\begin{equation}\label{eq:wi}
w(i) = M(\xi_i)v(i) + J(\xi_i)^T\bar{q}(i),
\end{equation}
where we reconstruct using $w(i)= 0$ if $\xi_i=0$, i.e., if all descriptions are lost at time $i$.\footnote{Instead of zero one can use the mean $\mathbb{E}[v]$ or the past value $v(i-1)$.} If $\xi_i\geq 1$ descriptions are received, we reconstruct using $w(i) = v(i) + q_{\xi_i}$. 

Recall that $v=z^{-1}L_w w + L_y y$, and let $(A_{yv}, B_{yv}, C_{yv}, D_{yv})$ and  $(A_{wv}, B_{wv}, C_{wv}, D_{wv})$ be minimal state-space representations for the transfer functions $L_y$ and $L_w$, respectively. Using this notation, we may express $v(i)$ in state-space form:
\begin{align}
x_{yv}(i+1) &= A_{yv} x_{yv}(i) + B_{yv} y(i) \\
v_y(i) &= C_{yv} x_{yv}(i) + D_{yv} y(i) \\
x_{wv}(i+1) &= A_{wv} x_{wv}(i) + B_{wv} w(i-1) \\
v_w(i) &= C_{wv} x_{wv}(i) + D_{wv} w(i-1) \\
v(i) &= v_y(i) + v_w(i).
\end{align}
Similarly, using \eqref{eq:system}, \eqref{eq:state_recursion}, \eqref{eq:yi}, and \eqref{eq:wi}, we get the following state-space expressions for $x(i)$ and $w(i)$:
\begin{align*}
x(i+1) &= 
 A x(i) + B_1d(i) + B_2FM(\xi)(C_{yv}x_{yv}(i)\\ \notag
&\quad + D_{yv}(C_2 x(i) + D_{21} d(i))
+ C_{wv} x_{wv}(i) \\
&\quad + D_{wv} w(i-1)) + B_2F J(\xi)^T\bar{q}(i),\\
w(i) & 
= M(\xi)(C_{yv} x_{yv}(i) + D_{yv} (C_2 x(i) + D_{21} d(i))\\
&\quad +C_{wv} x_{wv}(i) + D_{wv} w(i-1) ) +  J(\xi)^T\bar{q}(i). 
\end{align*}

%
Let us now form the augmented state-vector $\bar{x}(i)$:
\begin{align}\label{eq:barx}
\bar{x}(i) =
\begin{bmatrix}
x(i) \\
x_{yv}(i)\\
x_{wy}(i)\\
w(i-1) 
\end{bmatrix}.
\end{align}
Then $\{\bar{x}(i)\}$ and $\{ \xi_i\}$ are jointly Markov and  $\{\bar{x}(i), \xi_i\}$  forms a MJLS on the form~\eqref{eq:mjls},
where for $\xi_i = 1,\dotsc, k+1$, the $k+1$ switching matrices of the system are given by   \eqref{eq:A} and \eqref{eq:B}, where $\Sigma(\xi) = B_2 F M(\xi)$:
\begin{align}\label{eq:A}
&\mathcal{A}(\xi) = \\
&\begin{bmatrix}
A + \Sigma(\xi)D_{yv} C_2 & \Sigma(\xi)& \Sigma(\xi)C_{wv} & \Sigma(\xi) D_{wv} \\ \notag
B_{yv}C_2 & A_{yv} & 0 & 0 \\
0 & 0 & A_{wv} & B_{wv} \\
M(\xi)D_{yv} C_2 & M(\xi)C_{yv} & M(\xi) C_{wv} & M(\xi)D_{wv} 
\end{bmatrix}.
\end{align}

\begin{align}\label{eq:B}
\mathcal{B}(\xi) =
\begin{bmatrix}
B_1 + B_2 FM(\xi) D_{21} & B_2F J(\xi)^T \\
B_{yv} D_{21} & 0 \\
0 & 0 \\
M(\xi)D_{21} & J(\xi)^T 
\end{bmatrix}.
\end{align}

Since the packet losses are i.i.d., and the number of jump states is finite, it is easy to see that the sequence $\{\xi_i\}$ of jump states is stationary Markov and ergodic \cite{costa:2005}. 

Let $p_{j|j'} = \mathrm{Prob}( \xi_i = j | \xi_{i-1} = j')$ be the probability of being in state $\xi_{i-1}=j'$ at time $i-1$ and switching to state $\xi_i = j$ at time $i$. It follows that $p_{j|j'} = p_{j}$. At time $i$, we have that $p_j  = (1-p_l)^{\xi_i}p_l^{k-\xi_i}$, where $p_l$ is the packet loss probability and $\xi_i$ indicates the number of received packets at time $i$. 
Let the matrix $\mathbb{A}$ be given as:
\begin{align}\label{eq:AA}
&\mathbb{A} = \\  \notag
&
\begin{bmatrix}
p_{1|1}  \mathcal{A}(1) \otimes \mathcal{A}(1)  \   \cdots \ p_{1|k+1} \mathcal{A}(k+1) \otimes  \mathcal{A}(k+1) \\
p_{2|1}  \mathcal{A}(1) \otimes  \mathcal{A}(1)  \ \cdots  \ p_{2|k+1}\mathcal{A}(k+1) \otimes  \mathcal{A}(k+1) \\
\vdots \\
p_{k+1|1}  \mathcal{A}(1) \otimes  \mathcal{A}(1)\  \cdots \ p_{k+1|k+1}\mathcal{A}(k+1) \otimes  \mathcal{A}(k+1) 
\end{bmatrix}.
\end{align}
where $\otimes$ is the Kronecker matrix product. 
It was  shown in \cite{costa:2005}, that if the greatest absolute eigenvalue of $\mathbb{A}$ is strictly less than one, then the system in \eqref{eq:mjls} is MSS. Even though the system is time-varying since the decoder  depends upon the erasure pattern, the  system is still AWSS if it is MSS \cite{costa:2005}.

\end{proof}

\begin{proof}[\textbf{Corollary \ref{col:avgstb}}]
From the proof of Lemma~\ref{lem:mjls}, the quantization noise $q$ is an external signal of variance $\sigma_q^2$, which is  time varying since it depends upon the number of received descriptions. With probability $p_s(\ell)$, $\ell$ descriptions are received each of variance $\sigma^2(\ell)$. 
The average variance of the quantization noise is then given by $\bar{\sigma}^2 = \sum_{\ell = 0}^{k} p_s(\ell) \sigma^2(\ell)$.
 Since $\gamma$ defines the minimum SNR required for stability, we must have $\sigma_v^2\bar{\sigma}^{-2} > \gamma$. 
\end{proof}

\begin{IEEEbiography}[{\includegraphics[width=1in,height=1.25in,clip,keepaspectratio]{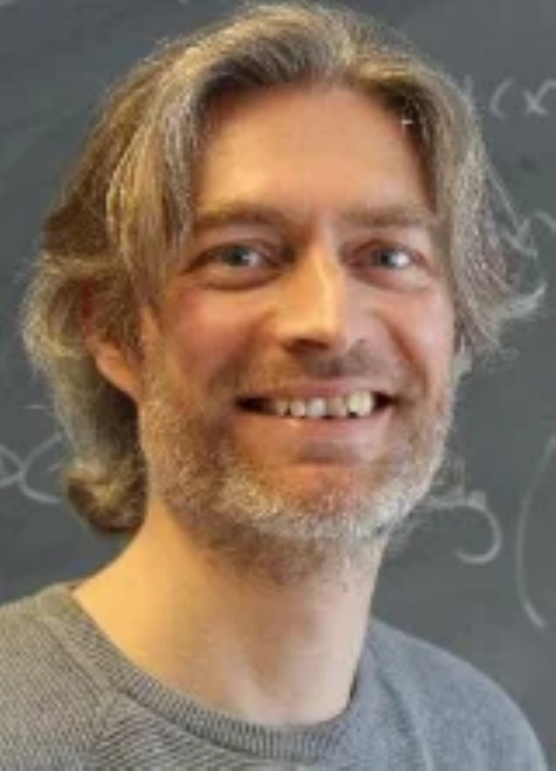}}]{\textbf{Jan {\O}stergaard (S’98-M’99-SM’11) received the M.Sc. degree from Aalborg University, Aalborg, Denmark, in 1999 and the PhD degree (with cum laude) from Delft University of Technology, Delft, The Netherlands, in 2007. From 1999 to 2002, he worked as an R\&D Engineer at ETI A/S, Aalborg, and from 2002 to 2003, he was an R\&D Engineer with ETI Inc., VA, USA. Between September 2007 and June 2008, he was a Postdoctoral Researcher at The University of Newcastle, NSW, Australia. He has been a Visiting Researcher at Tel Aviv University, Israel, and at Universidad Tecnica Federico Santa Maria, Valparaiso, Chile. Dr. Østergaard is currently a Full Professor in Information Theory and Signal Processing, Head of the Section on AI and Sound, and Head of the Centre on Acoustic Signal Processing Research (CASPR), at Aalborg University. He has received a Danish Independent Research Council’s Young Researcher’s Award, a Best PhD Thesis award by the European Association for Signal Processing (EURASIP), and fellowships from the Danish Independent Research Council and the Villum Foundations Young Investigator Programme. He is an Associate Editor of the IEEE Transactions on Information Theory.  His research interests are in the areas of statistical signal processing, information theory, source coding, joint source-channel coding, networked control theory, and acoustic signal processing.}}

\end{IEEEbiography}

\end{document}